\RequirePackage{etoolbox}
\csdef{input@path}{{style/}{graphics/}}
\documentclass[aap,MSNbibl,seceqn,dvips]{arximspdf}

\doi{10.1214/14-AAP1011} 
\volume{25}
\issue{2}
\pubyear{2015}
\firstpage{823}
\lastpage{859}

\makeatletter
\newcommand{\rrvert}{\vert}
\newcommand{\llvert}{\vert}
\newtheorem{theorem}{Theorem}[section]

\newtheorem{lemma}[theorem]{Lemma}
\newtheorem{lemmaa}{Lemma}[section]
\newproclaim{definition}[theorem]{Definition}
\newproclaim{remark}[theorem]{Remark}
\newproclaim{example}[theorem]{Example}

\newcommand{\eqref}[1]{(\ref{#1})}

\newcommand{\eps}{\varepsilon}

\renewcommand{\H}{\mathbb{H}}
\newcommand{\N}{\mathbb{N}}
\newcommand{\R}{\mathbb{R}}

\newcommand{\cA}{\mathcal{A}}
\newcommand{\cB}{\mathcal{B}}
\newcommand{\cC}{\mathcal{C}}
\newcommand{\cE}{\mathcal{E}}
\newcommand{\cF}{\mathcal{F}}
\newcommand{\cH}{\mathcal{H}}
\newcommand{\cI}{\mathcal{I}}
\newcommand{\cL}{\mathcal{L}}
\newcommand{\cP}{\mathcal{P}}
\newcommand{\cQ}{\mathcal{Q}}
\newcommand{\cR}{\mathcal{R}}
\newcommand{\cU}{\mathcal{U}}

\newcommand{\bA}{\mathbf{A}}

\newcommand{\fP}{\mathfrak{P}}

\newcommand{\graph}{\operatorname{graph}}
\newcommand{\proj}{\operatorname{proj}}
\newcommand{\ri}{\operatorname{ri}}
\newcommand{\NA}{\operatorname{NA}}
\newcommand{\USA}{\operatorname{USA}}
\newcommand{\supp}{\operatorname{supp}}
\newcommand{\linspan}{\operatorname{span}}
\newcommand{\Index}{\operatorname{index}}

\newcommand{\1}{\mathbf{1}}
\newcommand{\sint}{\stackrel{\mbox{\tiny$\bullet$}}{}}

\renewcommand{\emptyset}{\varnothing}

\makeatother

\begin{document}
\begin{frontmatter}

\title{Arbitrage and duality in nondominated discrete-time models}
\runtitle{Arbitrage and duality in nondominated models}

\begin{aug}
\author[A]{\fnms{Bruno}~\snm{Bouchard}\thanksref{T1}\ead[label=e1]{bouchard@ceremade.dauphine.fr}}
\and
\author[B]{\fnms{Marcel}~\snm{Nutz}\corref{}\thanksref{T2}\ead[label=e2]{mnutz@math.columbia.edu}}
\runauthor{B. Bouchard and M. Nutz}
\affiliation{CEREMADE,
Universit\'e Paris Dauphine
and CREST-ENSAE\\
 and Columbia University}
\address[A]{CEREMADE\\
Universit\'e Paris Dauphine\\
Place du Mar\'{e}chal de Lattre de Tassigny\\
75775 Paris Cedex 16\\
France\\
and\\
CREST-ENSAE\\
\printead{e1}} 
\address[B]{Department of Mathematics\\
Columbia University\\
2990 Broadway\\
New York, New York 10027\\
USA\\
\printead{e2}}
\end{aug}
\thankstext{T1}{Research supported by ANR Liquirisk and
Investissements d'Avenir (ANR-11-IDEX-0003/Labex Ecodec/ANR-11-LABX-0047).}
\thankstext{T2}{Research supported by NSF Grant DMS-1208985.}

\received{\smonth{6} \syear{2013}}

%
\begin{abstract}
We consider a nondominated model of a discrete-time financial market
where stocks
are traded dynamically, and options are available for static hedging.
In a general measure-theoretic setting, we show that absence of
arbitrage in a
quasi-sure sense is equivalent to the existence of a suitable family of
martingale
measures. In the arbitrage-free case, we show that optimal superhedging
strategies
exist for general contingent claims, and that the minimal superhedging
price is
given by the supremum over the martingale measures. Moreover, we obtain a
nondominated version of the Optional Decomposition Theorem.
\end{abstract}

%

\end{frontmatter}

\section{Introduction}\label{se:intro}

We consider a discrete-time financial market where\break stocks and,
possibly, options are available as hedging instruments. The market is
governed by a set $\cP$ of probability measures, not necessarily
equivalent, whose role is to determine which events are negligible
(polar), and which ones are relevant. We thus unify two approaches: On
the one hand, the framework of model uncertainty, where each $P\in\cP
$ is seen as a possible model for the stocks and a robust analysis over
the class $\cP$ is performed. On the other hand,
model-free approach, where no attempt is made to model the stocks
directly, but one sees the distribution of the stocks as partially
described by the current prices of the traded options---an ill-posed
inverse problem. Both approaches typically lead to a set $\cP$ which
is nondominated in the sense that there exists no reference probability
measure with respect to which all $P\in\cP$ are absolutely continuous.

We answer three fundamental questions of mathematical finance in this
context. The first one is how to formulate a condition of market
viability and relate it to the existence of consistent pricing
mechanisms. The condition $\NA(\cP)$ stated below postulates that
there is no trading strategy which yields a nonnegative gain that is
strictly positive on a relevant event; that is, an event that has
positive probability for at least one $P\in\cP$. We obtain a version
of the ``\hyperref[fft]{First Fundamental Theorem}'' stating that $\NA(\cP)$ holds if
and only if there exists a family $\cQ$ of martingale measures which
is equivalent to $\cP$ in the sense that $\cQ$ and $\cP$ have the
same polar sets. The next question is the one of superhedging: for a
contingent claim $f$, what is the minimal price $\pi(f)$ that allows
the seller to offset the risk of $f$ by trading appropriately in the
given stocks and options? We show in the ``\hyperref[st]{Superhedging Theorem}'' below
that $\pi(f)=\sup_{Q\in\cQ} E_Q[f]$, which corresponds to the
absence of a duality gap in the sense of convex analysis, and moreover
that an optimal superhedging strategy exists. From these two theorems,
it will follow that the precise range of arbitrage-free prices of $f$
is given by the open interval $(-\pi(-f),\pi(f))$ if $-\pi(-f)\neq
\pi(f)$, and by the singleton $\pi(f)$ otherwise. This latter case
occurs if and only if $f$ can be replicated (up to a polar set) by
trading in the hedging instruments. Finally, we obtain a version of the
``\hyperref[sft]{Second Fundamental Theorem},'' stating that the market is complete
(i.e., all claims are replicable) if and only if $\cQ$ is a singleton.

In addition to these main financial results, let us mention two
probabilistic conclusions that arise from our study. The first one is a
nondominated version of the Optional Decomposition Theorem, stating
that a process which is a supermartingale under all of the (not
necessarily equivalent) martingale measures $Q\in\cQ$ can be written
as the difference of a martingale transform and an increasing process.
Second, our theory can be used to prove an intriguing conjecture of
\cite{AcciaioBeiglbockPenknerSchachermayer.12}; namely, that every
martingale inequality (in finite discrete time) can be reduced to a
deterministic inequality in Euclidean space; cf.\ Remark~\ref{rk:martIneq}.

The main difficulty in our endeavor is that $\cP$ can be nondominated,
which leads to the failure of various tools of probability theory and
functional analysis; in particular, the Dominated Convergence Theorem,
Koml\'os' lemma (in the sense of \cite{DelbaenSchachermayer.94}) and
important parts of the theory of $L^p$ spaces. As a consequence, we
have not been able to reach general results by using separation
arguments in appropriate function spaces, which is the classical
approach of mathematical finance in the case where $\cP$ is a
singleton. Instead, we proceed in a ``local'' fashion; our basic
strategy is to first answer our questions in the case of a one-period
market with deterministic initial data and then ``glue'' the solutions
together to obtain the multi-period case. To solve the superhedging
problem in the one-period case, we show in a first step that (the
nontrivial inequality of) the duality holds for certain approximate
martingale measures, and in a second step, we use a strengthened
version of the \hyperref[fft]{First Fundamental Theorem} to show that a suitable
perturbation allows one to turn an approximate martingale measure into
a true one. To perform the gluing, we tailor our theory such as to be
compatible with the classical theory of analytic sets.

In the remainder of the \hyperref[se:intro]{Introduction}, we detail the notation, state the
main financial results and review the extant literature. In Section~\ref{se:existence}, we obtain the existence of optimal superhedging
strategies; this part is formulated in a more general setting than the
main results because we do not resort to a local analysis. In Section~\ref{se:onePeriod}, we prove the \hyperref[fft]{First Fundamental Theorem} and the
\hyperref[st]{Superhedging Theorem} in the one-period case. In Section~\ref{se:multiperiodWithoutOpt}, we obtain the same theorems in the
multi-period case, under the hypothesis that only stocks are traded. In
Section~\ref{se:multiperiodWithOpt}, we add the options to the picture
and prove the main results (as well as a slightly more precise form of
them). Section~\ref{se:optDecomp} discusses the Optional Decomposition Theorem, and the \hyperref[se:appendix]{Appendix} collects some results of
martingale theory that are used in the body of the paper.

\subsection{Notation}

Given a measurable space $(\Omega,\cA)$, we denote by $\fP(\Omega)$
the set of all probability measures on $\cA$. If $\Omega$ is a
topological space, $\cB(\Omega)$ denotes its Borel $\sigma$-field,
and we always endow $\fP(\Omega)$ with the topology of weak
convergence; in particular, $\fP(\Omega)$ is Polish whenever $\Omega
$ is Polish. The \emph{universal completion} of $\cA$ is the $\sigma
$-field $\bigcap_{P\in\fP(\Omega)} \cA^P$, where $\cA^P$ is the
$P$-completion of $\cA$. If $\Omega$ is Polish, $A\subseteq\Omega$
is \emph{analytic} if it is the image of a Borel subset of another
Polish space under a Borel-measurable mapping. A function $f\dvtx  \Omega
\to\overline{\R}:=[-\infty,\infty]$ is \emph{upper semianalytic}
if the super-level set $\{f>c\}$ is analytic for all $c\in\R$. Any
Borel set is analytic, and any analytic set is universally measurable
[i.e., measurable for the universal completion of $\cB(\Omega)$];
similarly, any Borel function is upper semianalytic, and any upper
semianalytic function is universally measurable. We refer to \cite
{BertsekasShreve.78}, Chapter~7, for these facts.
Given $P\in\fP(\Omega)$, we define the $P$-expectation for \emph
{any} measurable function $f\dvtx  \Omega\to\overline{\R}$ by
%
\begin{equation}
\label{eq:conventionExpectation} \hspace*{22pt}E_P[f]:= E_P\bigl[f^+\bigr]-E_P
\bigl[f^-\bigr]\qquad  \mbox{with the convention }\infty-\infty:= -\infty.
\end{equation}
The term \emph{random variable} is reserved for measurable functions
with values in $\R$. We shall often deal with a family $\cP\subseteq
\fP(\Omega)$ of measures. Then
a subset $A\subseteq\Omega$ is called \emph{$\cP$-polar} if
$A\subseteq A'$ for some $A'\in\cA$ satisfying $P(A')=0$ for all
$P\in\cP$, and a property is said to hold \emph{$\cP$-quasi surely}
or \emph{$\cP$-q.s. }if it holds outside a $\cP$-polar set. (In
accordance with this definition, any set is $\cP$-polar in the trivial
case $\cP=\emptyset$.)

\subsection{Main results}\label{se:introMainRes}

Let $T\in\N$ be the time horizon, and let $\Omega_1$ be a Polish
space. For $t\in\{0,1,\dots,T\}$, let $\Omega_t:=\Omega_1^t$ be the
$t$-fold Cartesian
product, with the convention that $\Omega_0$ is a singleton. We denote
by $\cF_t$ the universal completion of $\cB(\Omega_t)$ and write
$(\Omega,\cF)$ for $(\Omega_T,\cF_T)$. This will be our basic
measurable space and we shall often see $(\Omega_t,\cF_t)$ as a
subspace of $(\Omega,\cF)$.
For each $t\in\{0,1,\dots,T-1\}$ and $\omega\in\Omega_t$, we are
given a nonempty convex set
$\cP_t(\omega)\subseteq\fP(\Omega_1)$ of probability measures;
we think of $\cP_t(\omega)$ as the set of possible models for the
$t$th period, given state $\omega$ at time $t$. Intuitively, the role
of $\cP_t(\omega)$ is to determine which events at the ``node''
$(t,\omega)$ are negligible (polar), and which ones are relevant (have
positive probability under at least one model). We assume that for each $t$,
\[
\graph(\cP_t):=\bigl\{(\omega,P)\dvtx  \omega\in
\Omega_t, P\in\cP _t(\omega)\bigr\}\subseteq
\Omega_t \times\fP(\Omega_t)\qquad  \mbox{is analytic.}
\]
This ensures that $\cP_t$ admits a universally measurable selector;
that is, a universally measurable kernel $P_t\dvtx  \Omega_t\to\fP(\Omega
_1)$ such that $P_t(\omega)\in\cP_t(\omega)$ for all $\omega\in
\Omega_t$. If we are given such a kernel $P_t$ for each $t\in\{
0,1,\dots, T-1\}$, we can define a probability $P$ on $\Omega$ by
Fubini's theorem,
\begin{eqnarray*}
P(A)&=&\int_{\Omega_1} \cdots\int_{\Omega_1}
\1_A(\omega_1,\dots,\omega_T)\\
&&\hphantom{\int_{\Omega_1} \cdots\int_{\Omega_1}}{}\times
P_{T-1}(\omega_1,\dots,\omega_{T-1}; d
\omega_T)\cdots P_0(d\omega_1), \qquad A\in\Omega,
\end{eqnarray*}
where we write $\omega=(\omega_1,\dots,\omega_T)$ for a generic
element of $\Omega\equiv\Omega_1^T$. The above formula will be
abbreviated as $P=P_0\otimes P_1\otimes\cdots\otimes P_{T-1}$ in the
sequel. We can then introduce the set $\cP\subseteq\fP(\Omega)$ of
possible models for the multi-period market up to time $T$,
\[
\cP:=\bigl\{P_0\otimes P_1\otimes\cdots
\otimes P_{T-1}\dvtx  P_t(\cdot)\in \cP_t(\cdot),
t=0,1,\dots, T-1\bigr\},
\]
where, more precisely, each $P_t$ is a universally measurable selector
of $\cP_t$. Or equivalently, to state the same in reverse, $\cP$ is
the set of all $P\in\fP(\Omega)$ such that any decomposition
$P=P_0\otimes P_1\otimes\cdots\otimes P_{T-1}$ into kernels $P_t$
satisfies $P_t(\cdot)\in\cP_t(\cdot)$ up to a $(P_0\otimes
P_1\otimes\cdots\otimes P_{t-1})$-nullset.

Let $d\in\N$, and let $S_t=(S_t^1,\dots,S_t^d)\dvtx  \Omega_t \to\R^d$
be Borel-measurable for all $t\in\{0,1,\dots,T\}$. We think of $S_t$
as the (discounted) prices of $d$ traded \emph{stocks} at time $t$.
Moreover, let $\cH$ be the set of all predictable $\R^d$-valued
processes, the \emph{trading strategies}. Given $H\in\cH$, the
corresponding wealth process (from vanishing initial capital) is the
discrete-time integral
%
\begin{equation}
\label{eq:defSint} H\sint S=(H\sint S_t)_{t\in\{0,1,\dots,T\}},\qquad  H\sint
S_t=\sum_{u=1}^t
H_u \Delta S_u,
\end{equation}
where $\Delta S_u=S_u-S_{u-1}$ is the price increment, and $H_u \Delta
S_u$ is a shorthand for the inner product $\sum_{i=1}^d H^{i}_u \Delta
S^i_u$ on $\R^d$.

Moreover, let $e\in\N\cup\{0\}$, and let $g=(g^1,\dots,g^e)\dvtx  \Omega
\to\R^d$ be Borel-measurable. Each $g^i$ is seen as a traded \emph
{option} which can be bought or sold at time $t=0$ at the price
$g^i_0$; without loss of generality, $g^i_0=0$. Following \cite
{Dupire.94}, the options can be traded only statically (i.e., at
$t=0$), which accounts for the difference in liquidity and other market
parameters compared to stocks. Given a vector $h\in\R^e$, the value
of the corresponding option portfolio is then given by $hg=\sum_{i=1}^e h^ig^i$, and a pair $(H,h)\in\cH\times\R^e$ is called a
\emph{semistatic hedging strategy.}

As in any model of mathematical finance, a no-arbitrage condition is
needed for the model to be viable. The following formulation seems
natural in our setup.

%
\begin{definition}\label{def:NA}
Condition $\NA(\cP)$ holds if for all $(H,h)\in\cH\times\R^e$,
\[
H\sint S_T +hg\geq0 \qquad \cP\mbox{-q.s.} \quad \mbox{implies}\quad  H\sint
S_T +hg= 0 \qquad \cP\mbox{-q.s.}
\]
\end{definition}

We observe that $\NA(\cP)$ reduces to the classical condition ``$\NA
$'' in the case where $\cP$ is a singleton. More generally, if $\NA(\{
P\})$ holds for all $P\in\cP$, then $\NA(\cP)$ holds trivially, but
the converse is false.

The two extreme cases of our setting occur when $\cP$ consists of all
measures on $\Omega$ and when $\cP$ is a singleton.

%
\begin{example}\label{ex:canonicalSetup}
(i) Let $\Omega=(\R^{d})^T$ and let $S$ be the coordinate-mapping
process; more precisely, $S_0\in\R^d$ is fixed and $S_t(\omega
)=\omega_t$ for $\omega=(\omega_1,\dots,\omega_T)\in(\R^{d})^T$
and $t>0$. Moreover, let $\cP_t(\omega)=\fP(\R^d)$ for all
$(t,\omega)$; then $\cP$ is simply the collection of all probability
measures on $\Omega$. We emphasize that a $\cP$-q.s. inequality is in
fact a pointwise inequality in this setting, as $\cP$ contains all
Dirac measures. Consider the case without options ($e=0$); we show that
$\NA(\cP)$ holds. Indeed, suppose there exists $H\in\cH$ such that
$H\sint S_T(\omega)\geq0$ for all $\omega\in\Omega$ and $H\sint
S_T(\bar{\omega})>0$ for some $\bar{\omega}\in\Omega$. Then, if
we consider the smallest $t\in\{1,\dots,T\}$ such that $H\sint
S_t(\bar{\omega})>0$ and let $\bar{\omega}_t':=-\bar{\omega
}_t+2\bar{\omega}_{t-1}$, the path $\tilde{\omega}:=(\bar{\omega
}_1,\dots,\bar{\omega}_{t-1},\bar{\omega}_t',\dots,\bar{\omega
}_t')$ satisfies
\[
H\sint S_T(\tilde{\omega})= H\sint S_{t-1}(\bar{\omega}) -
H_t(\bar {\omega}) (\bar{\omega}_t - \bar{
\omega}_{t-1}) <0,
\]
a contradiction.

We note that in this canonical setup, a contingent claim is necessarily
a functional of $S$. We can generalize this setting by taking $\Omega
=(\R^{d'})^T$ for some $d'>d$ and defining $S$ to be the first $d$
coordinates of the coordinate-mapping process. Then, the remaining
$d'-d$ components play the role of nontradable assets, and we can model
claims that depend on additional risk factors.

(ii) Suppose we want to retrieve the classical case where we have a
single measure $P$. Given $P\in\fP(\Omega)$, the existence of
conditional probability distributions on Polish spaces \cite{StroockVaradhan.79}, Theorem~1.1.6, page 13, implies that there are Borel
kernels $P_t\dvtx  \Omega_t\to\fP(\Omega_1)$ such that $P=P_0\otimes
P_1\otimes\cdots\otimes P_{T-1}$, where $P_0$ is the restriction
$P|_{\Omega_1}$. If we then take $\cP_t(\omega):=\{P_t(\omega)\}$,
the Borel-measurability of $\omega\mapsto P_t(\omega)$ implies that
$\graph(\cP_t)$ is Borel for all $t$, and we have $\cP=\{P\}$ as desired.
\end{example}

Similarly as in the classical theory, the absence of arbitrage will be
related to the existence of linear pricing mechanisms. A probability
$Q\in\fP(\Omega)$ is a \emph{martingale measure} if $S$ is a
$Q$-martingale in the filtration $(\cF_t)$. We are interested in
martingale measures $Q$ which are absolutely continuous with respect to
$\cP$ in the sense that $Q(N)=0$ whenever $N$ is $\cP$-polar; it
seems natural to denote this relation by $Q\ll\cP$. Let us also
introduce a stronger notion of absolute continuity: we write
\[
Q\lll\cP \qquad \mbox{if there exists }P\in\cP\mbox{ such that }Q\ll P.
\]
We shall use this stronger notion, mainly because it is better suited
to the theory of analytic sets which will be used extensively later on.
(However, a posteriori, one can easily check that our main results hold
also with the weaker notion $Q\ll\cP$.)

If $Q$ is a martingale measure, then $Q$ is consistent with the given
prices of the traded options $g^i$ if $E_Q[g^i]=0$ for $i=1,\dots,e$.
We thus define the set
%
\begin{eqnarray}
\label{eq:defcQ} \cQ&=& \bigl\{Q\lll\cP\dvtx\nonumber\\[-8pt]\\[-8pt]
&&\hphantom{\bigl\{}  Q\mbox{ is a martingale measure and }
E_Q\bigl[g^i\bigr]=0\mbox{ for }i=1,\dots,e \bigr\}.\nonumber
\end{eqnarray}

Our first main result states that $\NA(\cP)$ is equivalent to $\cQ$
being sufficiently rich, in the sense that any $\cQ$-polar set is $\cP
$-polar. In this case, $\cQ$ is equivalent to $\cP$ in terms of polar
sets, which yields an appropriate substitute for the classical notion
of an equivalent martingale measure. In the body of this paper, we
shall work with the seemingly stronger condition (ii) below; it turns
out to be the most convenient formulation for our theory.

\begin{stFTAP*}\label{fft}
The following are equivalent:
\begin{longlist}[(ii$^\prime$)]
\item[(i)] $\NA(\cP)$ holds;
\item[(ii)]  for all $P\in\cP$ there exists $Q\in\cQ$ such that $P \ll Q$;
\item[(ii$^\prime$)] $\cP$ and $\cQ$ have the same polar sets.
\end{longlist}
\end{stFTAP*}

The following result, which is the main goal of our study, establishes
the dual characterization for the superhedging price $\pi(f)$ of a
contingent claim $f$ and the existence of an optimal superhedging
strategy. We use the standard convention $\inf\emptyset= +\infty$.

\begin{superhedThm*}\label{st}
Let $\NA(\cP)$ hold, and let $f\dvtx \Omega\to\R$ be upper
semianalytic. Then the minimal superhedging price
\[
\pi(f):=\inf \bigl\{x\in\R\dvtx  \exists (H,h)\in\cH\times\R^e \mbox {
such that } x+ H\sint S_T + hg\geq f\ \cP\mbox{-q.s.} \bigr\}
\]
satisfies
\[
\pi(f)=\sup_{Q\in\cQ} E_Q[f] \in(-\infty,\infty],
\]
and there exist $(H,h)\in\cH\times\R^e$ such that $\pi(f)+H\sint
S_T + hg\geq f\ \cP$-q.s.
\end{superhedThm*}

We remark that the supremum over $\cQ$ is not attained in general. For
instance, if $\cP=\fP(\Omega)$ while $S$ is constant and $e=0$, we
have $\sup_{Q\in\cQ} E_Q[f]=\sup_{\omega\in\Omega} f(\omega)$,
which clearly need not be attained. On the other hand, the supremum is
attained if, for instance, $f$ is a bounded continuous function, and
$\cQ$ is weakly compact.

%
\begin{remark}
The \hyperref[st]{Superhedging Theorem} also yields the following Lagrange duality: if
$\cQ'$ is the set of \emph{all} martingale measures $Q\lll\cP$ (not
necessarily consistent with the prices of the options $g^i$), then
\[
\pi(f)=\inf_{h\in\R^e}\sup_{Q\in\cQ'}
E_Q[f-hg].
\]
\end{remark}

\begin{remark}\label{rk:martIneq}
The \hyperref[st]{Superhedging Theorem} implies that every martingale inequality can
be derived from a deterministic inequality, in the following sense.
Consider the canonical setting of Example~\ref{ex:canonicalSetup}, and
let $f$ be as in the theorem. Then $\cQ$ is the set of all martingale
laws, and hence the inequality $E[f(M_1,\dots,M_T)]\leq a$ holds for
every martingale $M$ starting at $M_0=S_0$ if and only if $\sup_{Q\in
\cQ} E_Q[f]\leq a$. Thus the theorem implies that any martingale
inequality of the form $E[f(M_1,\dots,M_T)]\leq a$ can be deduced from
a completely deterministic (``pathwise'') inequality of the form $f\leq
a+ H\sint S_T$ by taking expectations. This fact is studied in more
detail in \cite{BeiglbockNutz.14}.
\end{remark}

A random variable $f$ is called \emph{replicable} if there exist $x\in
\R$ and $(H,h)\in\cH\times\R^e$ such that $x+H\sint S_T +hg = f\ \cP$-q.s. We say that the market is \emph{complete} if all
Borel-measurable functions $f\dvtx \Omega\to\R$ are replicable. Similarly
as in the classical theory, we have the following result.

\begin{ndFTAP*}\label{sft}
Let $\NA(\cP)$ hold and let $f\dvtx \Omega\to\R$ be upper semianalytic.
The following are equivalent:
\begin{longlist}[(iii)]
\item[(i)]$f$ is replicable;
\item[(ii)] the mapping $Q\mapsto E_Q[f]$ is constant (and finite) on $\cQ$;
\item[(iii)] for all $P\in\cP$ there exists $Q\in\cQ$ such that $P\ll Q$
and $E_Q[f]=\pi(f)$.
\end{longlist}
Moreover, the market is complete if and only if $\cQ$ is a singleton.
\end{ndFTAP*}

Regarding (iii), we remark that the existence of \emph{one} $Q\in\cQ
$ with $E_Q[f]=\pi(f)$ is not sufficient for $f$ to be replicable.
Regarding (ii), we observe that if the market is complete, the unique
element $Q\in\cQ$ satisfies $E_Q[f]\in\R$ for all real-valued Borel
functions $f$. This readily implies that $Q$ must be supported by a
finite set.
%

\subsection{Literature}

Our main financial results are, of course, extensions of the classical
case where $\cP$ is a singleton; see, for example, \cite
{DelbaenSchachermayer.06,FollmerSchied.04} and the references therein.
Our local way of proving the \hyperref[fft]{First Fundamental Theorem} is certainly
reminiscent of arguments for the Dalang--Morton--Willinger Theorem in
\cite{DalangMortonWillinger.90} and \cite{JacodShiryaev.98}. To the
best of our knowledge, our approach to the \hyperref[st]{Superhedging Theorem} is new,
even in the classical case.

In the setting where no options are traded, and $\cP$ is supposed to
consist of martingale measures in the first place, there is a
substantial literature on superhedging under ``volatility
uncertainty,'' where $S$ is a (continuous-time) process with continuous
trajectories; see, among others, \cite{BouchardMoreauNutz.12,FernholzKaratzas.11,MatoussiPiozinPossamai.12,MatoussiPossamaiZhou.12,NeufeldNutz.12,NutzSoner.10,NutzZhang.12,Peng.10,PossamaiRoyerTouzi.13,SonerTouziZhang.2010dual,SonerTouziZhang.2010rep,Song.10}.
These results were obtained by
techniques which do not apply in the presence of jumps. Indeed, an
important difference to our setting is that when $S$ is a continuous
(hence predictable) process, and each $P\in\cP$ corresponds to a
complete market, the optional decomposition coincides with the
Doob--Meyer decomposition, which allows for a constructive proof of the
\hyperref[st]{Superhedging Theorem}. In discrete time, a duality result (without
existence) for a specific topological setup was obtained in \cite
{DeparisMartini.04}; see also \cite{Dolinsky.13} for the case of
American and game options. A comparable result for the continuous case
was given in \cite{DenisMartini.06}. The existence of optimal
superhedging strategies in the discrete-time case was established in
\cite{Nutz.13} for the one-dimensional case $d=1$; its result is
generalized by our Theorem~\ref{th:optimalStrategy} to the
multidimensional case, with a much simpler proof using the arguments of
\cite{KabanovStricker.01note}. An abstract duality result was also
provided in \cite{Nutz.13}, but it remained open whether there is a
duality gap with respect to ($\sigma$-additive) martingale measures.

Roughly speaking, the model-free approach corresponds to the case where
$\cP$ consists of all probability measures on $\Omega$.
Starting with \cite{BrownHobsonRogers.01,CarrEllisGupta.98,Hobson.98},
one branch of this literature is devoted to finding explicitly a
model-independent semi-static hedging strategy for a specific claim,
such as a barrier option, when a specific family of options can be
traded, such as call options at specific maturities. We refer to the
survey \cite{Hobson.11} which features an extensive list of
references. Needless to say, explicit results are not within reach in
our general setting.

A more related branch of this literature, starting with \cite
{BreedenLitzenberger.78}, studies when a given set (finite or not) of
option prices is generated by a martingale measure and, more generally,
is consistent with absence of arbitrage; see \cite
{AcciaioBeiglbockPenknerSchachermayer.12,BiaginiCont.07,Buehler.06,CarrMadan.05,Cousot.07,CoxObloj.11,DavisHobson.07,Hobson.98}.
The recent paper \cite{AcciaioBeiglbockPenknerSchachermayer.12} is the
closest to our study. In the setting of Example~\ref
{ex:canonicalSetup} with $d=1$, it provides a version of the \hyperref[fft]{First Fundamental Theorem} and the absence of a duality gap, possibly with
infinitely many options, under certain continuity and compactness
conditions. The arguments hinge on the weak compactness of a certain
set of measures; to enforce the latter, it is assumed that options with
specific growth properties can be traded.

Adapting the terminology of the previous papers to our context, our
condition $\NA(\cP)$ excludes both model-independent and
model-dependent arbitrage. This is crucial for the validity of the
\hyperref[sft]{Second Fundamental Theorem} in a general context. For instance, it is
not hard to see that the theorem does not hold under the no-arbitrage
condition of \cite{AcciaioBeiglbockPenknerSchachermayer.12}; the
reason is that, in a typical case, the superhedging price $\pi(f)$
will be arbitrage-free for some models $P\in\cP$ and fail to be so
for others. On a related note, let us stress that the equivalence in
our version of the \hyperref[fft]{First Fundamental Theorem} provides not just one
martingale measure as in \cite
{AcciaioBeiglbockPenknerSchachermayer.12}, but a family $\cQ$
equivalent to $\cP$. This concept seems to be new, and is important
for our general version of the \hyperref[st]{Superhedging Theorem}. Finally, let us
remark that in general, optimal superhedging strategies do not exist
when infinitely many options are traded.

In the case where all options (or equivalently, all call options) at
one or more maturities are available for static hedging, consistency
with the option prices is equivalent to a constraint on the marginals
of the martingale measures. In this case, the problem dual to
superhedging is a so-called martingale optimal transport problem.
Following work of Beiglb{\"o}ck, Henry-Labord{\`e}re and Penkner \cite
{BeiglbockHenryLaborderePenkner.11} in discrete time and Galichon,
Henry-Labord{\`e}re and Touzi \cite{GalichonHenryLabordereTouzi.11} in
continuous time, this can be used to prove the absence of a duality gap
under certain regularity conditions, and in specific cases, to
characterize a dual optimizer and a superhedging strategy. See also
\cite{BeiglbockJuillet.12,DolinskySoner.12,DolinskySoner.13,HenryLabordereTouzi.13,TanTouzi.11} for recent developments.

The present study is also related to the theory of nonlinear
expectations, and we make use of ideas developed in \cite
{Nutz.10Gexp,NutzVanHandel.12} in that context. In particular, the
dynamic version of the superhedging price used in the proof of the
\hyperref[st]{Superhedging Theorem} in Section~\ref{se:multiperiodWithoutOpt} takes
the form of a conditional sublinear expectation. In that section, we
also make heavy use of the theory of analytic sets and related
measurable selections; see \cite{Wagner.77,Wagner.80} for an extensive survey.

\section{Existence of optimal superhedging strategies}\label{se:existence}

In this section, we obtain the existence of optimal superhedging
strategies via an elementary closedness property with respect to
pointwise convergence. The technical restrictions imposed on the
structure of $\Omega,\cP$ and $S$ in the \hyperref[se:intro]{Introduction} are not
necessary for this, and so we shall work in a more general setting: For
the remainder of this section, $\cP$ is any nonempty collection of
probability measures on a general measurable space $(\Omega,\cF)$
with filtration $\{\cF_t\}_{t\in\{0,1,\dots,T\}}$, and
$S=(S_0,S_1,\dots, S_T)$ is any collection of $\cF$-measurable, $\R
^d$-valued random variables $S_t$. To wit, the process $S$ is \emph
{not necessarily adapted}. The purpose of this unusual setup is the following.

\begin{remark}
Static hedging with options can be incorporated as a special case of a
non-adapted stock: Suppose we want to model dynamic trading in
(typically adapted) stocks $S^1,\dots,S^d$ as well as static trading
in $\cF$-measurable random variables (options) $g^1,\dots,g^e$ at
initial prices $g^i_0,\dots,g^e_0$ ($\cF$-measurable at least, but
typically $\cF_0$-measurable). Then we can define the $d+e$
dimensional process $\tilde{S}$ by $\tilde{S}_t^i= S^i_t$ for $i\leq
d$ and
\[
\tilde{S}_0^{i+d}= g^i_0 \quad \mbox{and}\quad
\tilde{S}_t^{i+d}= g^i, \qquad t=1,\dots,T
\]
for $i=1,\dots, e$.
Since $\Delta\tilde{S}_1^{i+d}=g^i-g^i_0$ and $\Delta\tilde
{S}_t^{i+d}=0$ for $t\geq2$, dynamic trading in $\tilde{S}$ is then
equivalent to dynamic trading in $S$ and static ($\cF_0$-measurable)
trading in $g^1,\dots,g^e$.
\end{remark}

In view of the previous remark, we do not have to consider the case
with options explicitly; that is, we take $e=0$. The only role of the
filtration is to determine the set $\cH$ of trading strategies; as in
the \hyperref[se:intro]{Introduction}, this will be the set of all predictable processes.
(The arguments in this section could easily be extended to situations
such as portfolio constraints.) For $H\in\cH$, the wealth process
$H\sint S$ is defined as in \eqref{eq:defSint}, and the condition $\NA
(\cP)$ says that
$H\sint S_T\geq0\ \cP$-q.s. implies $H\sint S_T= 0\ \cP$-q.s.

We write $\cL^0_+$ for the set of all nonnegative random variables.
The following result states that the cone $\cC$ of all claims which
can be superreplicated from initial capital $x=0$ is closed under
pointwise convergence.

\begin{theorem}\label{th:closedness}
Let $\cC:=\{H\sint S_T\dvtx  H\in\cH\} - \cL^0_+$. If $\NA(\cP)$
holds, then $\cC$ is closed under $\cP$-q.s. convergence; that is, if
$\{W^n\}_{n\geq1} \subseteq\cC$ and $W$ is a random variable such
that $W^n\to W\ \cP$-q.s., then $W\in\cC$.
\end{theorem}

Before stating the proof, let us show how this theorem implies the
existence of optimal superreplicating strategies.
%

\begin{theorem}\label{th:optimalStrategy}
Let $\NA(\cP)$ hold, and let $f$ be a random variable. Then
\[
\pi(f):=\inf \{x\in\R\dvtx  \exists H\in\cH\mbox{ such that } x+ H\sint
S_T\geq f\ \cP\mbox{-q.s.} \}>-\infty,
\]
and there exists $H\in\cH$ such that $\pi(f) + H\sint S_T \geq f\ \cP$-q.s.
\end{theorem}

\begin{pf}
The claim is trivial if $\pi(f)=\infty$. Suppose that $\pi
(f)=-\infty$. Then, for all $n\geq1$, there exists $H^n\in\cH$ such that
$-n+H^n\sint S_T \geq f\ \cP$-q.s. and hence
\[
H^n\sint S_T \geq f+n \geq(f+n)\wedge1\qquad  \cP\mbox{-q.s.}
\]
That is, $W^n:=(f+n)\wedge1\in\cC$ for all $n\geq1$. Now Theorem~\ref{th:closedness} yields that $1=\lim W^n$ is in $\cC$, which
clearly contradicts $\NA(\cP)$.

On the other hand, if $\pi(f)$ is finite, then $W^n:=f-\pi(f)-1/n \in
\cC$ for all $n\geq1$ and thus $f-\pi(f) = \lim W^n \in\cC$ by
Theorem~\ref{th:closedness}, which yields the existence of $H$.
\end{pf}

\begin{pf*}{Proof of Theorem~\ref{th:closedness}}
We follow quite closely the arguments from \cite
{KabanovStricker.01note}; as observed in \cite{KabanovStricker.06},
they can be used even in the case where $S$ is not adapted. Let
\[
W^n= H^n\sint S_T - K^n
\]
be a sequence in $\cC$ which converges $\cP$-q.s. to a random
variable $W$; we need to show that $W=H\sint S_T - K$ for some $H\in
\cH$ and $K\in\cL^0_+$.
We shall use an induction over the number of periods in the market. The
claim is trivial when there are zero periods. Hence, we show the
passage from $T-1$ to $T$ periods; more precisely, we shall assume that
the claim is proved for any market with dates $\{1,2,\dots,T\}$, and
we deduce the case with dates $\{0,1,\dots,T\}$.

For any real matrix $M$, let $\Index(M)$ be the number of rows in $M$
which vanish identically. Now let
$\H_1$ be the random $(d\times\infty)$-matrix whose columns are
given by the vectors $H_1^1,H_1^2,\dots\,$. Then $\Index(\H_1)$ is a
random variable with values in $\{0,1,\dots,d\}$.
If $\Index(\H_1)=d\ \cP$-q.s., we have $H^n_1=0$ for all $n$, so
that setting $H_1=0$, we conclude immediately by the induction
assumption. For the general case, we use another induction over
$i=d,d-1,\dots,0$; namely, we assume that the result is proved
whenever $\Index(\H_1)\geq i\ \cP$-q.s., and we show how to pass to $i-1$.

Indeed, assume that $\Index(\H_1)\geq i-1\in\{0,\dots,d-1\}$; we
shall construct $H$ separately on finitely many sets forming a
partition of $\Omega$. Consider first the set
\[
\Omega_1:=\bigl\{\liminf\bigl|H_1^n\bigr|<\infty\bigr
\} \in\cF_0.
\]
By a standard argument (e.g., \cite{KabanovStricker.01note}, Lemma~2),
we can find $\cF_0$-measurable random indices $n_k$ such that on
$\Omega_1$, $H_1^{n_k}$ converges pointwise to a (finite) $\cF
_0$-measurable random vector $H_1$. As the sequence
\[
\tilde{W}^k:=W^{n_k}-H^{n_k}_1 \Delta
S_1 = \sum_{t=2}^T
H^{n_k}_t \Delta S_t - K^{n_k}
\]
converges to $W-H_1\Delta S_1=:\tilde{W}\ \cP$-q.s. on $\Omega_1$,
we can now apply the induction assumption to obtain $H_2,\dots, H_T$
and $K\geq0$ such that
\[
\tilde{W} = \sum_{t=2}^T H_t
\Delta S_t - K
\]
and therefore $W=H\sint S_T-K$ on $\Omega_1$.
It remains to construct $H$ on
\[
\Omega_2:=\Omega_1^c=\bigl\{
\liminf\bigl|H_1^n\bigr|=+\infty\bigr\}.
\]
Let
\[
G_1^n:= \frac{H_1^n}{1+|H_1^n|}.
\]
As $|G_1^n|\leq1$, there exist $\cF_0$-measurable random indices
$n_k$ such that $G_1^{n_k}$ converges pointwise to an $\cF
_0$-measurable random vector $G_1$, and clearly $|G_1|=1$ on $\Omega_2$.
Moreover, on $\Omega_2$, we have $W^{n_k}/(1+|H_1^{n_k}|) \to0$ and
hence $- G_1\Delta S_1$ is the $\cP$-q.s. limit of
\[
\sum_{t=2}^T \frac{H^{n_k}_t}{{1+|H_1^{n_k}|}} \Delta
S_t - \frac
{K^{n_k}}{{1+|H_1^{n_k}|}}.
\]
By the induction assumption, it follows that there exist $\tilde
{H}_2,\dots, \tilde{H}_T$ such that $\sum_{t=2}^T \tilde{H}_t\Delta
S_t\ge- G_1\Delta S_1$ on $\Omega_{2}\in\cF_{0}$. Therefore,
%
\begin{equation}
\label{eq:Gnonneg} G_1\Delta S_1+\sum
_{t=2}^T \tilde{H}_t\Delta
S_t=0 \qquad \mbox{on } \Omega_{2},
\end{equation}
since otherwise the trading strategy $(G_{1},\tilde{H}_{2},\ldots,\tilde{H}_{T})\1_{\Omega_{2}}$ would violate\break $\NA(\cP)$.
As $|G_1|=1$ on $\Omega_2$, we have that for every $\omega\in\Omega
_2$, at least one component $G_1^j(\omega)$ of $G_1(\omega)$ is
nonzero. Therefore,
\begin{eqnarray*}
\Lambda_1&:=&\Omega_2 \cap\bigl\{G_1^1
\neq0\bigr\},\\
  \Lambda_j&:=&\bigl(\Omega_2 \cap \bigl
\{G_1^j\neq0\bigr\}\bigr) \setminus(\Lambda_1
\cup\cdots\cup\Lambda_{j-1}),\qquad  j=2,\dots,d
\end{eqnarray*}
defines an $\cF_0$-measurable partition of $\Omega_2$. We then
consider the vectors
\[
\bar{H}^n_t:=H^n_t - \sum
_{j=1}^d \1_{\Lambda_j}
\frac
{H^{n,j}_1}{G^j_1} ( G_1 \1_{\{t=1\}}+ \tilde{H}_{t}
\1_{\{t\ge
2\} } ),\qquad  t=1,\ldots,T.
\]
Note that $\bar{H}^n \sint S_T = H^n \sint S_T$ by \eqref
{eq:Gnonneg}. Hence, we still have
$W=\bar{H}^n \sint S_T-K^n$. However, on $\Omega_2$, the resulting
matrix $\bar{\H}_1$
now has $\Index(\bar{\H}_1)\geq i$ since we have created an
additional vanishing row: the $j$th component
of $\bar{H}^n_1$ vanishes on $\Lambda_j$ by construction, while the
$j$th row of $\H_1$
cannot vanish on $\Lambda_j\subseteq\{G_1^j\neq0\}$ by the
definition of $G_1$. We can now apply the induction hypothesis for
indices greater or equal to $i$ to obtain $H$ on $\Omega_2$. Recalling
that $\Omega=\Omega_1\cup\Omega_2$, we have shown that there exist
$H\in\cH$ and $K\geq0$ such that $W=H\sint S_T - K$.
\end{pf*}

\section{The one-period case}\label{se:onePeriod}

In this section, we prove the \hyperref[fft]{First Fundamental} \hyperref[fft]{Theorem} and the
\hyperref[st]{Superhedging Theorem} in the one-period case; these results will serve
as building blocks for the multi-period case. We consider an arbitrary
measurable space $(\Omega,\cF)$ with a filtration $(\cF_0,\cF_1)$,
where $\cF_0=\{\emptyset,\Omega\}$, and a nonempty convex set $\cP
\subseteq\fP(\Omega)$. The stock price process is given by a
deterministic vector $S_0\in\R^d$ and an $\cF_1$-measurable, $\R
^d$-valued random vector $S_1$. We write $\Delta S$ for $S_1-S_0$ and
note that $Q$ is a martingale measure simply if $E_Q[\Delta S]=0$;
recall the convention \eqref{eq:conventionExpectation}. Moreover, we
have $\cH=\R^d$. We do not consider hedging with options explicitly
(i.e., we again take $e=0$) as there is no difference between options
and stocks in the one-period case. Thus we have $\cQ=\{Q\in\fP
(\Omega)\dvtx  Q\lll\cP, E_Q[\Delta S]=0\}$.

\subsection{First fundamental theorem}

In the following version of the fundamental theorem, we omit the
equivalent condition (ii$^\prime$) stated in the\break \hyperref[se:intro]{Introduction}; it is not
essential, and we shall get back to it only in Remark~\ref
{rk:proofsForMainRes}.

\begin{theorem}\label{th:FTAP1}
The following are equivalent:
\begin{longlist}[(ii)]
\item[(i)]$\NA(\cP)$ holds;
\item[(ii)] for all $P\in\cP$ there exists $Q\in\cQ$ such that $P \ll Q$.
\end{longlist}
\end{theorem}

For the proof, we first recall the following well-known fact; cf.\ \cite
{DellacherieMeyer.82}, Theorem VII.57, page 246.

\begin{lemma}\label{le:integrableProbab}
Let $\{g_n\}_{n\geq1}$ be a sequence of random variables, and let
$P\in\fP(\Omega)$. There exists a probability measure $R\sim P$ with
bounded density $dR/dP$ such that $E_R[|g_n|]<\infty$ for all $n$.
\end{lemma}

The following lemma is a strengthening of the nontrivial implication in
Theorem~\ref{th:FTAP1}. Rather than just showing the existence of a
martingale measure (i.e., $E_R[\Delta S]=0$ for some $R\in\fP(\Omega
)$), we show that the vectors $E_R[\Delta S]$ fill a relative
neighborhood of the origin; this property will be of key importance in
the proof of the duality. We write $\ri A$ for the relative interior of
a set $A$.

\begin{lemma}[(Fundamental lemma)]\label{le:relInt}
Let $\NA(\cP)$ hold, and let $f$ be a random variable. Then
\[
0\in\ri \bigl\{ E_R[\Delta S]\dvtx  R\in\fP(\Omega), R \lll\cP,
E_R\bigl[|\Delta S|+|f|\bigr]<\infty \bigr\}\subseteq\R^d.
\]
Similarly, given $P\in\cP$, we also have
\[
0\in\ri \bigl\{ E_R[\Delta S]\dvtx  R\in\fP(\Omega), P \ll R \lll\cP,
E_R\bigl[|\Delta S|+|f|\bigr]<\infty \bigr\}\subseteq\R^d.
\]
\end{lemma}

\begin{pf}
We show only the second claim; the first one can be obtained by
omitting the lower bound $P$ in the subsequent argument. We fix $P$ and
$f$; moreover, we set $\cI_{k}:=\{1,\ldots,d\}^{k}$ for $k=1,\dots,d$ and
\[
\Theta:=\bigl\{R\in\fP(\Omega)\dvtx  P \ll R \lll\cP, E_R\bigl[|\Delta
S|+|f|\bigr]<\infty\bigr\}.
\]
Note that $\Theta\neq\emptyset$ as a consequence of Lemma~\ref
{le:integrableProbab}. Given $I\in\cI_{k}$, we denote
\[
\Gamma_{I}:=\bigl\{ E_{R}\bigl[\bigl(\Delta
S^{i}\bigr)_{i\in I}\bigr]\dvtx  R\in\Theta\bigr\}\subseteq
\R^k;
\]
then our claim is that $0\in\ri\Gamma_{I}$ for $I=(1,\dots,d)\in
\cI_d$. It is convenient to show more generally that $0\in\ri\Gamma
_{I}$ for all $I\in\cI_k$ and all $k=1,\dots,d$. We proceed by induction.

Consider first $k=1$ and $I\in\cI_k$; then $I$ consists of a single
number $i\in\{1,\dots,d\}$. If $\Gamma_{I}=\{0\}$, the result holds
trivially, so we suppose that there exists $R\in\Theta$ such that
$E_{R}[\Delta S^{i}] \neq0$. We focus on the case $E_{R}[\Delta S^{i}]
>0$; the reverse case is similar. Then, $\NA(\cP)$ implies that $A:=\{
\Delta S^{i}<0\}$ satisfies $R_1(A)>0$ for some $R_1\in\cP$. By
replacing $R_1$ with $R_2:=(R_1+P)/2$ we also have that $R_2\gg P$, and
finally Lemma~\ref{le:integrableProbab} allows to replace $R_2$ with
an equivalent probability $R_3$ such that $E_{R_3}[|\Delta
S|+|f|]<\infty$; as a result, we have found $R_3\in\Theta$
satisfying $E_{R_3}[\1_{A}\Delta S^{i}]<0$. But then $R'\sim R_3\gg P$
defined by
%
\begin{equation}
\label{eq:densityArgumentForRelInt} \frac{dR'}{dR_3}= \frac{\1_A + \eps}{E_{R_3}[\1_A + \eps]}
\end{equation}
satisfies $R'\in\Theta$ and $E_{R'}[\Delta S^{i}]<0$ for $\eps>0$
chosen small enough.
Now set
\[
R_\lambda:= \lambda R + (1-\lambda)R' \in\Theta
\]
for each $\lambda\in(0,1)$; then
\[
0\in\bigl\{E_{R_\lambda}\bigl[\Delta S^i\bigr]\dvtx  \lambda\in(0,1)
\bigr\} \subseteq\ri\bigl\{ E_{R}\bigl[\Delta S^i\bigr]\dvtx  R
\in\Theta\bigr\},
\]
which was the claim for $k=1$.

Let $1<k\leq d$ be such that $0\in\ri\Gamma_{I}$ for all $I\in\cI
_{k-1}$; we show hat $0\in\ri\Gamma_{I}$ for all $I\in\cI_{k}$.
Suppose that there exists $I=(i_1,\dots,i_k)\in\cI_{k}$ such that
$0\notin\ri\Gamma_{I}$. Then, the convex set $\Gamma_{I}$ can be
separated from the origin; that is, we can find $y=(y^{1},\ldots,y^{k})\in\R^{k}$ such that $|y|=1$ and
\[
0\leq\inf \Biggl\{E_{R} \Biggl[\sum_{j=1}^{k}
y^{j} \Delta S^{i_j} \Biggr]\dvtx  R\in\Theta \Biggr\}.
\]
Using an argument similar to that which precedes \eqref
{eq:densityArgumentForRelInt}, this implies that $\sum_{j=1}^{k} y^{j}
\Delta S^{i_j}\ge0\ \cP$-q.s., and thus $\sum_{j=1}^{k} y^{j}
\Delta S^{i_j}=0\ \cP$-q.s. by $\NA(\cP)$. As $|y|=1$, there exists
$1\leq l\leq k$ such that $y^l\ne0$, and we obtain that
\[
\Delta S^{l}= -\sum_{j=1}^{k}
\delta_{j\ne l} \bigl(y^{j} /y^l\bigr) \Delta
S^{i_j} \qquad \cP\mbox{-q.s.}
\]
Using the definition of the relative interior, the assumption that
$0\notin\ri\Gamma_{I}$ then implies that $0\notin\ri\Gamma_{I'}$,
where $I'\in\cI_{k-1}$ is the vector obtained from $I$ by deleting
the $l$th entry. This contradicts our induction hypothesis.
\end{pf}

\begin{pf*}{Proof of Theorem~\ref{th:FTAP1}}
(i) implies (ii): This is a special case of Lemma~\ref{le:relInt},
applied with $f\equiv0$.

(ii) implies (i): Let (ii) hold, and let $H\in\R^d$ be such that
$H\Delta S\geq0\ \cP$-q.s.
Suppose that there exists $P\in\cP$ such that $P\{H\Delta S >0\}>0$.
By (ii), there exists a martingale measure $Q$ such that $P \ll Q \lll
\cP$. Thus $Q\{H\Delta S >0\}>0$, contradicting that $E_Q[H\Delta S]=0$.
\end{pf*}

\subsection{Superhedging theorem}

We can now establish the \hyperref[st]{Superhedging}\break \hyperref[st]{Theorem} in the one-period case.
Recall that convention \eqref{eq:conventionExpectation} is in force.

\begin{theorem}\label{th:duality1}
Let $\NA(\cP)$ hold, and let $f$ be a random variable. Then
%
\begin{eqnarray}
\label{eq:dualityOnePeriod} \sup_{Q\in\cQ} E_Q[f] = \pi(f)
:=\inf
\bigl\{x\in\R\dvtx  \exists H\in \R^d\mbox{ such that } x+ H\Delta S\geq f
\ \cP\mbox{-q.s.} \bigr\}.\nonumber
\\
\end{eqnarray}
Moreover, $\pi(f)>-\infty$, and there exists $H\in\R^d$ such that
$\pi(f)+H\Delta S \geq f\ \cP$-q.s.
\end{theorem}

The last statement is a consequence of Theorem~\ref
{th:optimalStrategy}. For the proof of \eqref{eq:dualityOnePeriod},
the inequality ``$\geq$'' is the nontrivial one; that is, we need to
find $Q_n\in\cQ$ such that $E_{Q_n}[f]\to\pi(f)$. Our construction
proceeds in two steps. In the subsequent lemma, we find ``approximate''
martingale measures $R_n$ such that $E_{R_n}[f]\to\pi(f)$; in its
proof, it is important to relax the martingale property as this allows
us to use arbitrary measure changes. In the second step, we replace
$R_n$ by true martingale measures, on the strength of the Fundamental Lemma: it implies that if $R$ is any probability with $E_{R}[\Delta S]$
close to the origin, then there exists a perturbation of $R$ which is
a martingale measure.

\begin{lemma}\label{le:approxOptimalMartMeas}
Let $\NA(\cP)$ hold, and let $f$ be a random variable with $\pi
(f)=0$. There exist probabilities $R_n\lll\cP$, $n\geq1$ such that
\[
E_{R_n}[\Delta S]\to0 \quad \mbox{and}\quad  E_{R_n}[f]\to0.
\]
\end{lemma}

\begin{pf}
It follows from Lemma~\ref{le:integrableProbab} that the set
\[
\Theta:=\bigl\{R\in\fP(\Omega)\dvtx  R \lll\cP, E_R\bigl[|\Delta S|+|f|\bigr]<
\infty \bigr\}
\]
is nonempty. Introduce the set
\[
\Gamma:= \bigl\{ E_R\bigl[(\Delta S, f)\bigr]\dvtx  P\in\Theta \bigr\}
\subseteq\R^{d+1};
\]
then our claim is equivalent to $0\in\overline{\Gamma}$, where
$\overline{\Gamma}$ denotes the closure of $\Gamma$ in $\R^{d+1}$.
Suppose for contradiction that $0\notin\overline{\Gamma}$, and note that
$\Gamma$ is convex because $\cP$ is convex. Thus, $\overline{\Gamma
}$ can be separated strictly from the origin; that is, there exist
$(y,z)\in\R^d\times\R$ with $|(y,z)|=1$ and $\alpha>0$ such that
\[
0 < \alpha= \inf_{R\in\Theta} E_R[y\Delta S+ zf].
\]
Using again a similar argument as before \eqref
{eq:densityArgumentForRelInt}, this implies that
%
\begin{equation}
\label{eq:proofApproxOptMartMeas} 0 < \alpha\leq y\Delta S+ zf \qquad \cP\mbox{-q.s.}
\end{equation}
Suppose that $z<0$; then this yields that
\[
f \leq\bigl|z^{-1}\bigr|y\Delta S - \bigl|z^{-1}\bigr|\alpha\qquad  \cP\mbox{-q.s.},
\]
which implies that $\pi(f) \leq-|z^{-1}|\alpha<0$ and thus
contradicts the assumption that $\pi(f)=0$. Hence, we must have $0\leq
z \leq1$. But as $\pi(zf)=z\pi(f)=0<\alpha/2$, there exists $H\in
\R^d$ such that $\alpha/2 + H\Delta S\geq zf\ \cP$-q.s., and then
\eqref{eq:proofApproxOptMartMeas} yields
\[
0< \alpha/2 \leq(y+H)\Delta S\qquad  \cP\mbox{-q.s.},
\]
which contradicts $\NA(\cP)$. This completes the proof.
\end{pf}

\begin{lemma}\label{le:martingaleCorrector}
Let $\NA(\cP)$ hold, let $f$ be a random variable and let $R\in\fP
(\Omega)$ be such that $R\lll\cP$ and $E_{R}[|\Delta S|+|f|]<\infty
$. Then there exists $Q\in\cQ$ such that $E_Q[|f|]<\infty$ and
\[
\bigl|E_Q[f]-E_R[f]\bigr| \leq c \bigl(1+ \bigl|E_{R}[f]\bigr|
\bigr)\bigl|E_{R}[\Delta S]\bigr|,
\]
where $c>0$ is a constant independent of $R$ and $Q$.
\end{lemma}

\begin{pf}
Let $\Theta=\{R'\in\fP(\Omega)\dvtx  R' \lll\cP, E_{R'}[|\Delta
S|+|f|]<\infty\}$ and
\[
\Gamma=\bigl\{E_{R'}[\Delta S]\dvtx  R'\in\Theta\bigr\}.
\]
If $\Gamma=\{0\}$, then $R\in\Theta$ is itself a martingale measure,
and we are done. So let us assume that the vector space $\linspan
\Gamma\subseteq\R^d$ has dimension $k>0$, and let $e_1,\dots,e_k$
be an orthonormal basis. Lemma~\ref{le:relInt} shows that
$0\in\ri\Gamma$; hence, we can find $P^{\pm}_{i} \in\Theta$ and
$\alpha^{\pm}_{i}>0$ such that
\[
\alpha^{\pm}_{i} E_{P^{\pm}_{i}}[\Delta S]=\pm
e_{i},\qquad  1\leq i\leq k.
\]
Note also that $P^{\pm}_{i}, \alpha^{\pm}_{i}$ do not depend on $R$.

Let $\lambda=(\lambda_1,\dots,\lambda_k)\in\R^k$ be such that $-
E_{R}[\Delta S]=\sum_{i=1}^{k}\lambda_{i}e_{i}$. Then we have
$|\lambda|=|E_{R}[\Delta S]|$ and
\[
-E_{R}[\Delta S]=\int\Delta S \,d\mu\qquad  \mbox{for } \mu:= \sum
_{i=1}^{k} \lambda_i^+
\alpha^{+}_{i} P^{+}_{i} +
\lambda_i^-\alpha^{-}_{i} P^{-}_{i},
\]
where $\lambda_i^+$ and $\lambda_i^-$ denote the positive and the
negative part of $\lambda_i$.
Define the probability $Q$ by
\[
Q = \frac{R+\mu}{1+\mu(\Omega)};
\]
then $R\ll Q\lll\cP$ and $E_{Q}[\Delta S]=0$ by construction. Moreover,
\begin{eqnarray*}
\bigl|E_Q[f]-E_R[f]\bigr| &=& \biggl\llvert \frac{1}{1+\mu(\Omega)}\int
f \,d\mu - \frac{\mu(\Omega
)}{1+\mu(\Omega)} E_R[f]\biggr\rrvert
\\
&\leq&\biggl\llvert \int f \,d\mu\biggr\rrvert + \mu(\Omega) \bigl\llvert
E_R[f]\bigr\rrvert
\\
&\leq& c|\lambda| \bigl(1+ \bigl|E_R[f]\bigr|\bigr),
\end{eqnarray*}
where $c$ is a constant depending only on $\alpha^{\pm}_{i}$ and
$E_{P^{\pm}_{i}}[f]$. It remains to recall that $|\lambda
|=|E_{R}[\Delta S]|$.
\end{pf}

\begin{pf*}{Proof of Theorem~\ref{th:duality1}}
The last claim holds by Theorem~\ref{th:optimalStrategy}, so $\pi
(f)>-\infty$. Let us first assume that $f$ is bounded from above; then
$\pi(f)<\infty$, and by a translation we may even suppose that $\pi(f)=0$.
By Theorem~\ref{th:FTAP1}, the set $\cQ$ of martingale measures is
nonempty; moreover, $E_Q[f]\leq\pi(f)=0$ for all $Q\in\cQ$ by Lemma~\ref{le:easyIneq}. Thus, we only need to find a sequence $Q_n\in\cQ$
such that $E_{Q_n}[f]\to0$. Indeed, Lemma~\ref
{le:approxOptimalMartMeas} yields a sequence $R_n\lll\cP$ such that
$E_{R_n}[\Delta S]\to0$ and $E_{R_n}[f]\to0$. Applying Lemma~\ref
{le:martingaleCorrector} to each $R_n$, we obtain a sequence $Q_n\in
\cQ$ such that
$E_{Q_n}[|f|]<\infty$ and
\[
\bigl|E_{Q_n}[f]-E_{R_n}[f]\bigr| \leq c \bigl(1+
\bigl|E_{R_n}[f]\bigr|\bigr)\bigl|E_{R_n}[\Delta S]\bigr|\to0;
\]
as a result, we have $E_{Q_n}[f]\to0$ as desired.

It remains to discuss the case where $f$ is not bounded from above. By
the previous argument, we have
%
\begin{equation}
\label{eq:dualityProofReductionBddAbove} \sup_{Q\in\cQ}E_Q[f\wedge n] = \pi(f
\wedge n), \qquad n\in\N;
\end{equation}
we pass to the limit on both sides. Indeed, on the one hand, we have
\[
\sup_{Q\in\cQ}E_Q[f\wedge n] \nearrow\sup
_{Q\in\cQ}E_Q[f]
\]
by the monotone convergence theorem (applied to all $Q$ such that
$E_Q[f^-]<\infty$). On the other hand, it also holds that
\[
\pi(f\wedge n)\nearrow\pi(f),
\]
because if $\alpha:=\sup_n \pi(f\wedge n)$, then $(f\wedge n) -
\alpha\in\cC$ for all $n$ and thus $f - \alpha\in\cC$ by Theorem~\ref{th:closedness}, and in particular $\pi(f)\leq\alpha$.
\end{pf*}

\section{The multi-period case without options}\label
{se:multiperiodWithoutOpt}

In this section, we establish the \hyperref[fft]{First Fundamental Theorem} and the
\hyperref[st]{Superhedging Theorem} in the market with $T$ periods, in the case where
only stocks are traded.

\subsection{Preliminaries on quasi-sure supports}

We first fix some terminology. Let $(\Omega,\cF)$ be any measurable
space, and let $Y$ be a topological space. A mapping $\Psi$ from
$\Omega$ into the power set of $Y$ will be denoted by $\Psi\dvtx  \Omega
\twoheadrightarrow Y$ and called a random set or a set-valued mapping.
We say that $\Psi$ is measurable (in the sense of set-valued mappings) if
%
\begin{equation}
\label{eq:defMblSetValued} \bigl\{\omega\in\Omega\dvtx  \Psi(\omega)\cap A\neq\emptyset\bigr\}\in
\cF \qquad \mbox{for all closed $A\subseteq Y$.}
\end{equation}
It is called closed-valued if $\Psi(\omega)\subseteq Y$ is closed for
all $\omega\in\Omega$.

\begin{remark}\label{rk:SetMeasurabilitySpecialCases}
The mapping $\Psi$ is called weakly measurable if
%
\begin{equation}
\label{eq:defWeaklyMblSetValued} \bigl\{\omega\in\Omega\dvtx  \Psi(\omega)\cap O\neq\emptyset\bigr\}\in
\cF \qquad \mbox{for all open $O\subseteq Y$.}
\end{equation}
This condition is indeed weaker than \eqref{eq:defMblSetValued},
whenever $Y$ is metrizable \cite{AliprantisBorder.06}, Lemma~18.2, page
593. If $Y=\R^d$, then \eqref
{eq:defWeaklyMblSetValued} is equivalent to \eqref
{eq:defMblSetValued}; cf.\ \cite{Rockafellar.76}, Proposition~1A.
\end{remark}

Another useful notion of measurability refers to the graph of $\Psi$,
defined as
\[
\graph(\Psi)=\bigl\{(\omega,y)\dvtx  \omega\in\Omega, y\in\Psi(\omega)\bigr\}
\subseteq\Omega\times Y.
\]
In particular, if $\Omega$ and $Y$ are Polish spaces and $\graph(\Psi
)\subseteq\Omega\times Y$ is analytic, then $\Psi$ admits a
universally measurable selector $\psi$ on the (universally measurable)
set $\{\Psi\neq\emptyset\}\subseteq\Omega$; that is, $\psi\dvtx  \{
\Psi\neq\emptyset\}\to Y$ satisfies $\psi(\omega)\in\Psi(\omega
)$ for all $\omega$ such that $\Psi(\omega)\neq\emptyset$.

\begin{lemma}
Given a nonempty family $\cR$ of probability measures on $(\R^d,\allowbreak \cB
(\R^d))$, let
\[
\supp(\cR):= \bigcap \bigl\{A\subseteq\R^d \mbox{ closed}\dvtx
R(A)=1\mbox{ for all }R\in\cR \bigr\} \subseteq \R^d.
\]
Then $\supp(\cR)$ is the smallest closed set $A\subseteq\R^d$ such
that $R(A)=1$ for all $R\in\cR$.
In particular, if $\cP\subseteq\fP(\Omega)$, $X\dvtx  \Omega\to\R^d$
is measurable, and $\cR=\{P\circ X^{-1}\dvtx  P\in\cP\}$ is the
associated family of laws, then
\[
\supp_{\cP}(X):= \supp(\cR) %
\]
is the smallest closed set $A\subseteq\R^d$ such that $P\{X\in A\}=1$
for all $P\in\cP$.
\end{lemma}

\begin{pf}
Let $\{B_n\}_{n\geq1}$ be a countable basis of the topology of $\R^d$
and let $A_n=B_n^c$. Since the intersection
\[
\supp(\cR)= \bigcap\bigl\{A_n\dvtx  R(A_n)=1
\mbox{ for all }R\in\cR\bigr\}
\]
is countable, we see that $R(\supp(\cR))=1$ for all $R\in\cR$. The
rest is clear.
\end{pf}

\begin{lemma}\label{le:supportMbl}
Let $\Omega, \Omega_1$ be Polish spaces, let $\cP\dvtx  \Omega
\twoheadrightarrow\fP(\Omega_1)$ have nonempty values and analytic
graph, and let $X\dvtx  \Omega\times\Omega_1\to\R^d$ be Borel.
Then the random set $\Lambda\dvtx  \Omega\twoheadrightarrow\R^d$,
\[
\Lambda(\omega):= \supp_{\cP(\omega)}\bigl(X(\omega,\cdot)\bigr)\subseteq
\R^d,\qquad  \omega\in\Omega
\]
is closed-valued and universally measurable. Moreover, its polar cone
$\Lambda^\circ\dvtx  \Omega\twoheadrightarrow\R^d$,
\[
\Lambda^\circ(\omega):=\bigl\{y\in\R^d\dvtx  yv\geq0\mbox{ for
all }v\in \Lambda(\omega)\bigr\},\qquad  \omega\in\Omega
\]
is nonempty-closed-valued and universally measurable, and it satisfies
%
\begin{equation}
\label{eq:polarConeQs} \Lambda^\circ(\omega) = \bigl\{y\in\R^d\dvtx  yX(
\omega,\cdot)\geq0\ \cP (\omega)\mbox{-q.s.}\bigr\},\qquad  \omega\in\Omega.
\end{equation}
\end{lemma}

\begin{pf}
Consider the mapping $\ell$ which associates to $\omega\in\Omega$
and $P\in\fP(\Omega_1)$ the law of $X(\omega,\cdot)$ under $P$,
\[
\ell\dvtx  \Omega\times\fP(\Omega_1) \to\fP\bigl(\R^d\bigr),\qquad
\ell(\omega,P):= P\circ X(\omega,\cdot)^{-1}.
\]
If $X$ is continuous and bounded, it is elementary to check that $\ell
$ is separately continuous [using the weak convergence on $\fP(\Omega
_1)$, as always] and thus Borel. A~monotone class argument then shows
that $\ell$ is Borel whenever $X$ is Borel. Consider also the random
set consisting of the laws of $X(\omega,\cdot)$,
\[
\cR\dvtx  \Omega\twoheadrightarrow\fP\bigl(\R^d\bigr),\qquad  \cR(\omega):= \ell
\bigl(\omega,\cP(\omega)\bigr)\equiv\bigl\{P\circ X(\omega,\cdot)^{-1}\dvtx
P\in\cP (\omega)\bigr\}.
\]
Its graph is the image of $\graph(\cP)$ under the Borel mapping
$(\omega,P)\mapsto(\omega,\ell(\omega,\allowbreak P))$;
in particular, $\graph(\cR)$ is again analytic by \cite{BertsekasShreve.78}, Proposition~7.40, page 165. Now let $O\subseteq\R^d$ be an
open set; then
\begin{eqnarray*}
\bigl\{\omega\in\Omega\dvtx  \Lambda(\omega) \cap O\neq\emptyset\bigr\} &=& \bigl\{
\omega\in\Omega\dvtx  R(O)>0 \mbox{ for some }R\in\cR(\omega)\bigr\}
\\
&=& \proj_{\Omega} \bigl\{(\omega,R)\in\graph(\cR)\dvtx  R(O)>0\bigr\},
\end{eqnarray*}
where $\proj_{\Omega}$ denotes the canonical projection $\Omega
\times\fP(\R^d)\to\Omega$. Since $R\mapsto R(O)$ is semicontinuous
and in particular Borel, this shows that
$\{\omega\in\Omega\dvtx\break  \Lambda(\omega) \cap O\neq\emptyset\}$ is
the continuous image of an analytic set, thus analytic and in
particular universally measurable.
In view of Remark~\ref{rk:SetMeasurabilitySpecialCases}, it follows
that $\Lambda$ is universally measurable as claimed.
Moreover, $\Lambda$ is closed-valued by its definition. Finally, the
polar cone $\Lambda^\circ$ is then also universally measurable \cite
{Rockafellar.76}, Proposition~1H, Corollary~2T, and it is clear that
$\Lambda^\circ$ is closed-valued and contains the origin.

It remains to prove \eqref{eq:polarConeQs}. Let $\omega\in\Omega$.
Clearly, $y\in\Lambda^\circ(\omega)$ implies that
\[
P\bigl\{yX(\omega,\cdot)\geq0\bigr\}=P\bigl\{X(\omega,\cdot)\in\Lambda(\omega )
\bigr\}=1
\]
for all $P\in\cP(\omega)$, so $y$ is contained in the right-hand
side of \eqref{eq:polarConeQs}. Conversely, let $y\notin\Lambda
^\circ(\omega)$; then there exists $v\in\Lambda(\omega)$ such that
$yv<0$, and thus $yv'<0$ for all $v'$ in an open neighborhood $B(v)$ of
$v$. By the minimality property of the support $\Lambda(\omega)$, it
follows that there exists $P\in\cP(\omega)$ such that $P\{X(\omega,\cdot)\in B(v)\} >0$. Therefore, $P\{yX(\omega,\cdot)<0\}>0$ and
$y$ is not contained in the right-hand side of \eqref{eq:polarConeQs}.
\end{pf}

%
\begin{remark}
The assertion of Lemma~\ref{le:supportMbl} cannot be generalized to
the case where $X$ is universally measurable instead of Borel (which
would be quite handy in Section~\ref{se:superhedMultiperiod} below).
Indeed, if $\Omega=\Omega_1=[0,1]$, $\cP\equiv\fP(\Omega)$ and
$X=\1_A$ for some $A\subseteq\Omega\times\Omega_1$, then $\{\Lambda
\cap\{1\}\neq\emptyset\}=\proj_\Omega(A)$. But there exists a
universally measurable subset $A$ of $[0,1]\times[0,1]$ whose
projection is not universally measurable; cf.\ the proof of \cite
{Fremlin.03}, 439G.
\end{remark}

\subsection{First fundamental theorem}

We now consider the setup as detailed in Section~\ref{se:introMainRes}, for the case when there are no options ($e=0$). In
brief, $(\Omega,\cF)=(\Omega_T,\cF_T)$, where $\Omega_t=\Omega
_1^t$ is the $t$-fold product of a Polish space $\Omega_1$ (and often
identified with a subset of $\Omega$) and $\cF_T$ is the universal
completion of $\cB(\Omega_T)$. The set $\cP$ is determined by the
random sets $\cP_t(\cdot)$, which have analytic graphs, and $S_t$ is
$\cB(\Omega_t)$-measurable. It will be convenient to write $S_{t+1}$
as a function on $\Omega_t\times\Omega_1$,
\[
S_{t+1}\bigl(\omega,\omega'\bigr) = S_{t+1}
\bigl(\omega_1,\dots,\omega_t,\omega '
\bigr), \qquad \bigl(\omega,\omega'\bigr)=\bigl((\omega_1,
\dots,\omega_t),\omega'\bigr)\in \Omega_t
\times\Omega_1,
\]
since the random variable $\Delta S_{t+1}(\omega,\cdot) =
S_{t+1}(\omega,\cdot) - S_{t+1}(\omega)$ on $\Omega_1$ will be of
particular interest. Indeed, $\Delta S_{t+1}(\omega,\cdot)$
determines a one-period market on $(\Omega_1,\cB(\Omega_1))$ under
the set $\cP_t(\omega)\subseteq\fP(\Omega_1)$. Endowed with
deterministic initial data, this market is of the type considered in
Section~\ref{se:onePeriod}, and we shall write $\NA(\cP_t(\omega))$
for the no-arbitrage condition in that market.
The following result contains the \hyperref[fft]{First Fundamental Theorem} for the
multi-period market, and also shows that absence of arbitrage in the
multi-period market is equivalent to absence of arbitrage in all the
one-period markets, up to a polar set.

%
\begin{theorem}\label{th:FTAP}
The following are equivalent:
\begin{longlist}[(iii)]
\item[(i)]$\NA(\cP)$ holds;
\item[(ii)] 
$N_t=\{\omega\in\Omega_t\dvtx  \NA(\cP_t(\omega)) \mbox
{ fails}\}$ is $\cP$-polar for all $t\in\{0,\dots,T-1\}$;
\item[(iii)] for all $P\in\cP$ there exists $Q\in\cQ$ such that $P \ll Q$.
\end{longlist}
\end{theorem}

Before stating the proof, let us isolate some preliminary steps.

\begin{lemma}\label{le:NAlocal}
Let $t\in\{0,\dots, T-1\}$. The set
\[
N_t=\bigl\{\omega\in\Omega_t\dvtx  \NA\bigl(
\cP_t(\omega)\bigr) \mbox{ fails}\bigr\}
\]
is universally measurable, and if $\NA(\cP)$ holds, then $N_t$ is
$\cP$-polar.
\end{lemma}

\begin{pf}
We fix $t$ and set $X^\omega(\cdot):=S_{t+1}(\omega,\cdot
)-S_t(\omega)$ for $\omega\in\Omega_t$. Let $\Lambda(\omega
)=\supp_{\cP_t(\omega)}(X^\omega)$, and let $\Lambda^\circ(\omega
)$ be its polar cone. Our first claim is that
%
\begin{equation}
\label{eq:NAandSupport} N_t^c\equiv\bigl\{\omega\dvtx  \NA\bigl(
\cP_t(\omega)\bigr)\mbox{ holds}\bigr\}= \bigl\{\omega \dvtx
\Lambda^\circ(\omega) = -\Lambda^\circ(\omega)\bigr\}.
\end{equation}
Indeed, suppose that $\Lambda^\circ(\omega) = -\Lambda^\circ
(\omega)$; then \eqref{eq:polarConeQs} shows that $yX^\omega\geq0\ \cP_t(\omega)$-q.s. implies $-yX^\omega\geq0\ \cP_t(\omega
)$-q.s., and hence $\NA(\cP_t(\omega))$ holds. On the other hand, if
there exists $y\in\Lambda^\circ(\omega)$ such that $-y\notin
\Lambda^\circ(\omega)$, then we have $y X^\omega\geq0\ \cP
_t(\omega)$-q.s. while $\{yX^\omega>0\}$ is not $\cP_t(\omega
)$-polar, meaning that $\NA(\cP_t(\omega))$ is violated. Therefore,
\eqref{eq:NAandSupport} holds.

Since $\Lambda^\circ(\cdot)$ is universally measurable by Lemma~\ref
{le:supportMbl}, it follows from the representation \eqref
{eq:NAandSupport} (and, e.g., \cite{Rockafellar.76}, Proposition~1A)
that $N_t$ is universally measurable.

It remains to show that $N_t$ is $\cP$-polar under $\NA(\cP)$.
Suppose for contradiction that there exists $P_*\in\cP$ such that
$P_*(N_t)>0$; then we need to construct (measurable) arbitrage
strategies $y(\omega)\in\R^d$ and measures $P(\omega)\in\cP
_t(\omega)$ under which $y(\omega)$ makes a riskless profit with
positive probability, for $\omega\in N_t$. In other words, we need to
select $P_*$-a.s. from
\[
\bigl\{(y,P)\in\Lambda^\circ(\omega)\times\cP_t(\omega)\dvtx
E_P\bigl[yX^\omega\bigr]>0\bigr\},\qquad  \omega\in N_t.
\]
To this end, note that by modifying as in \cite{BertsekasShreve.78}, Lemma~7.27, page
173 each member of a universally measurable
Castaing representation \cite{AliprantisBorder.06}, Corollary 18.14, page
601 of $\Lambda^\circ$, we can find a
Borel-measurable mapping $\Lambda^\circ_{*}\dvtx  \Omega_t
\twoheadrightarrow\R^d$ with nonempty closed values such that
%
\begin{equation}
\label{eq:LambdaStar} \Lambda^\circ_{*} = \Lambda^\circ\qquad  P_*
\mbox{-a.s.}
\end{equation}
This implies that $\graph(\Lambda^\circ_{*}) \subseteq\Omega
_t\times\R^d$ is Borel \cite{AliprantisBorder.06}, Theorem~18.6, page
596. Let
\[
\Phi(\omega):=\bigl\{(y,P)\in\Lambda^\circ_{*}(\omega)\times
\cP _t(\omega)\dvtx  E_P\bigl[yX^\omega\bigr]>0\bigr\},\qquad
\omega\in\Omega_t.
\]
After using a monotone class argument to see that the function
\[
\psi\dvtx  \Omega\times\fP(\Omega_1)\times\R^d\to\R, \qquad \psi(
\omega,P,y):= E_P\bigl[yX^\omega\bigr]
\]
is Borel, we deduce that (with a minor abuse of notation)
\[
\graph(\Phi)= \bigl[\graph(\cP_t) \times\R^d\bigr] \cap
\bigl[\fP(\Omega _1)\times\graph\bigl(\Lambda^\circ_{*}
\bigr)\bigr] \cap\{\psi>0\}
\]
is an analytic set. Now, we can apply the Jankov--von Neumann Theorem
\cite{BertsekasShreve.78}, Proposition~7.49, page 182 to obtain a
universally measurable function $\omega\mapsto(y(\omega),P(\omega))$
such that $(y(\cdot),P(\cdot))\in\Phi(\cdot)$ on $\{\Phi\neq
\emptyset\}$. By the definition of $N_t$ and \eqref{eq:LambdaStar},
we have
\[
N_t=\{\Phi\neq\emptyset\}\qquad  P_*\mbox{-a.s.},
\]
so that $y$ is $P_*$-a.s. an arbitrage on $N_t$. On the universally
measurable $P_*$-nullset $\{y \notin\Lambda^\circ\}$, we may
redefine $y:=0$ to ensure that $y$ takes values in $\Lambda^\circ$.
Similarly, on $\{\Phi= \emptyset\}$, we may define $P$ to be any
universally measurable selector of $\cP_t$. Setting $H_{t+1}:=y$ and
$H_s:=0$ for $s\neq t+1$, we have thus defined $H\in\cH$ and $P(\cdot
)\in\cP_t(\cdot)$ such that
%
\begin{equation}
\label{eq:HalmostArbitrage} P(\omega)\bigl\{H_{t+1}(\omega)X^\omega>0\bigr
\}>0\qquad  \mbox{for $P_*$-a.e. } \omega\in N_t
\end{equation}
and $H_{t+1}(\omega)X^\omega\geq0\ \cP_t(\omega)$-q.s. for all
$\omega\in\Omega$; cf.\ \eqref{eq:polarConeQs}.
As any $P'\in\cP$ satisfies a decomposition of the form $P'|_{\Omega
_{t+1}}=P'|_{\Omega_t}\otimes P'_t$ for some selector $P'_t$ of $\cP
_t$, Fubini's theorem yields that
$H\sint S_T\geq0\ \cP$-q.s. On the other hand, let
\[
P^*:=P_*|_{\Omega_t}\otimes P \otimes\tilde{P}_{t+1}\otimes\cdots
\otimes\tilde{P}_{T-1},
\]
where $\tilde{P}_s$ is any universally measurable selector of $\cP
_s$, for $s=t+1,\dots,T-1$. Then $P^*\in\cP$, while $P_*(N_t)>0$ and
\eqref{eq:HalmostArbitrage} imply that
$P^*\{H\sint S_T>0\}>0$. This contradicts $\NA(\cP)$ and completes
the proof that $N_t$ is $\cP$-polar.
\end{pf}

We recall the following result of Doob; it based on the fact that the
Borel $\sigma$-field of $\Omega_1$ is countably generated; see \cite
{DellacherieMeyer.82}, Theorem V.58, page 52 and the subsequent remark.

\begin{lemma}\label{le:RadonNikodymMbl}
Let $\mu, \mu'\dvtx  \Omega\times\fP(\Omega_1)\to\fP(\Omega_1)$ be
Borel. There exists a Borel function $D\dvtx  \Omega\times\Omega\times
\fP(\Omega_1) \times\fP(\Omega_1)\times\Omega_1 \to\overline
{\R}$ such that
\[
D\bigl(\omega,\omega',P,P';\tilde{\omega}\bigr) =
\frac{d\mu(\omega,P)}{d\mu'(\omega',P')} \biggl|_{\cB(\Omega_1)}(\tilde{\omega})
\]
for all $\omega,\omega'\in\Omega$ and $P,P'\in\fP(\Omega_1)$;
that is, $D(\omega,\omega',P,P';\cdot)$ is a version of the
Radon--Nikodym derivative of the absolutely continuous part of $\mu
(\omega,P)$ with respect to $\mu'(\omega',P')$, on the Borel $\sigma
$-field of $\Omega_1$.
\end{lemma}

%
\begin{lemma}\label{le:martMeasSelection}
Let $t\in\{0,\dots,T-1\}$, let $P(\cdot): \Omega_t\to\fP(\Omega
_1)$ be Borel, and let
\[
\cQ_t(\omega):=\bigl\{Q\in\fP(\Omega_1)\dvtx  Q\lll
\cP_t(\omega), E_Q\bigl[\Delta S_{t+1}(\omega,
\cdot)\bigr]=0\bigr\}, \qquad \omega\in\Omega_t.
\]
Then $\cQ_t$ has an analytic graph and there exist universally
measurable mappings $Q(\cdot),\hat{P}(\cdot): \Omega_t\to\fP
(\Omega_1)$ such that
\begin{eqnarray*}
P(\omega)&\ll& Q(\omega)\ll\hat{P}(\omega)\qquad  \mbox{for all }\omega \in
\Omega_t,
\\
\hat{P}(\omega)&\in&\cP_t(\omega) \qquad \mbox{if }P(\omega)\in\cP
_t(\omega),
\\
Q(\omega)&\in&\cQ_t(\omega)\qquad  \mbox{if }\NA\bigl(\cP_t(
\omega)\bigr)\mbox{ holds and }P(\omega)\in\cP_t(\omega).
\end{eqnarray*}
\end{lemma}

\begin{pf}
Given $\omega\in\Omega_t$, we write $X^\omega$ for $\Delta
S_{t+1}(\omega,\cdot)$. Let $P(\cdot)\dvtx  \Omega_t\to\fP(\Omega_1)$
be Borel. As a first step, we show that the random set
%
\begin{eqnarray}
\label{eq:tildeXii} \Xi(\omega)&:=&\bigl\{(Q,\hat{P})\in\fP(\Omega_1)\times
\fP(\Omega_1)\dvtx \nonumber\\[-8pt]\\[-8pt]
&&\hphantom{\bigl\{} E_Q\bigl[X^\omega\bigr]=0, \hat{P}
\in\cP_t(\omega), P(\omega)\ll Q\ll\hat {P}\bigr\}\nonumber
\end{eqnarray}
has an analytic graph. To this end, let
\[
\Psi(\omega):=\bigl\{Q\in\fP(\Omega_1)\dvtx  E_Q
\bigl[X^\omega\bigr]=0\bigr\}, \qquad \omega \in\Omega_t.
\]
Since the function
\[
\psi\dvtx  \Omega_t\times\fP(\Omega_1)\to\R,\qquad  \psi(\omega,R):=
E_R\bigl[X^\omega\bigr]
\]
is Borel, we see that
\[
\graph(\Psi)= \{\psi=0\} \subseteq\Omega_t\times\fP(
\Omega_1) \qquad \mbox{is Borel.}
\]
Next, consider the random set
\[
\Phi(\omega):=\bigl\{(R,\hat{R})\in\fP(\Omega_1)\times\fP(\Omega
_1)\dvtx  P(\omega) \ll R\ll\hat{R}\bigr\},\qquad  \omega\in\Omega_t.
\]
Moreover, let
\begin{eqnarray*}
\phi\dvtx  \Omega_t\times\fP(\Omega_1)\times\fP(
\Omega_1)&\to&\R, \\
\phi(\omega,R,\hat{R})&:=& E_{P(\omega)}\bigl[dR/dP(
\omega)\bigr]+ E_{R}[d\hat{R}/dR],
\end{eqnarray*}
where $dR/dP(\omega)$ and $d\hat{R}/dR$ are the jointly
Borel-measurable Radon--Niko\-dym derivatives as in Lemma~\ref
{le:RadonNikodymMbl}. Then $P(\omega)\ll R\ll\hat{R}$ if and only if
$\phi(\omega,R,\hat{R})=2$. Since Lemma~\ref{le:RadonNikodymMbl}
and \cite{BertsekasShreve.78}, Proposition~7.26, page 134, imply that
$\phi$ is Borel, we deduce that
\[
\graph(\Phi)= \{\phi=2\} \subseteq\Omega_t\times\fP(
\Omega_1)\qquad  \mbox{is Borel.}
\]
As a result, we obtain that
\[
\Xi(\omega) = \bigl[\Psi(\omega)\times\cP_t(\omega)\bigr] \cap\Phi (
\omega),\qquad  \omega\in\Omega_t
\]
has an analytic graph. Thus, the Jankov--von Neumann Theorem \cite
{BertsekasShreve.78}, Proposition~7.49, page 182, allows us to find
universally measurable selectors $Q(\cdot),\hat{P}(\cdot)$ for $\Xi
$ on the universally measurable set $\{\Xi\neq\emptyset\}$. Setting
$Q(\cdot):=P(\cdot)$ and $\hat{P}(\cdot):=P(\cdot)$ on $\{\Xi\neq
\emptyset\}$ completes the construction, as by Theorem~\ref
{th:FTAP1}, $\Xi(\omega)=\emptyset$ is possible only if $\NA(\cP
_t(\omega))$ does not hold or $P(\omega)\notin\cP_t(\omega)$.

It remains to show that $\cQ_t$ has an analytic graph. Using the same
arguments as for the measurability of $\Xi$, but omitting the lower
bound $P(\cdot)$, we see that the random set
%
\begin{eqnarray}
\label{eq:tildeXi} \tilde{\Xi}(\omega)&:=&\bigl\{(Q,\hat{P})\in\fP(
\Omega_1)\times\fP (\Omega_1)\dvtx \nonumber\\[-8pt]\\[-8pt]
&&\hphantom{\bigl\{} E_Q\bigl[
\Delta S_{t+1}(\omega,\cdot)\bigr]=0, \hat{P}\in\cP _t(
\omega), Q\ll\hat{P}\bigr\}\nonumber
\end{eqnarray}
has an analytic graph. Moreover, we observe that $\graph(\cQ_t)$ is
the image of $\graph(\tilde{\Xi})$ under the canonical projection
$\Omega_t\times\fP(\Omega_1)\times\fP(\Omega_1) \to\Omega
_t\times\fP(\Omega_1)$, so that $\graph(\cQ_t)$ is indeed analytic.
\end{pf}

\begin{pf*}{Proof of Theorem~\ref{th:FTAP}}
Lemma~\ref{le:NAlocal} shows that (i) implies (ii). Let (iii) hold;
then a set is $\cP$-polar if and only if it is $\cQ$-polar. Let $H\in
\cH$ be such that $H\sint S_T\geq0\ \cP$-q.s. For all $Q\in\cQ$,
we have $H\sint S_T\geq0\ Q$-a.s., and as Lemma~\ref
{le:genMartingales} shows that the local $Q$-martingale $H\sint S$ is a
true one, it follows that $H\sint S_T=0\ Q$-a.s. Hence, (i) holds.

It remains to show that (ii) implies (iii). Let $P\in\cP$; then
$P=P_0\otimes\cdots\otimes P_{T-1}$ for some universally measurable
selectors $P_t(\cdot)$ of $\cP_t(\cdot)$, $t=0,\dots,T-1$. We first
focus on $t=0$. Using Theorem~\ref{th:FTAP1}, we can find $\hat
{P}_0\in\cP_0$ and $Q_0\in\cQ_0$ such that $P_0\ll Q_0\ll\hat
{P}_0$. Next, consider $t=1$. By changing $P_1(\cdot)$ on a $\hat
{P}_0$-nullset (hence a $P_0$-nullset), we find a Borel kernel which
takes values in $\cP_1(\cdot)\ P_0$-a.e.; we again denote this
kernel by $P_1(\cdot)$. As any $\hat{P}_0$-nullset is a
$P_0$-nullset, this change does not affect the identity $P=P_0\otimes
\cdots\otimes P_{T-1}$. In view of (ii), we can apply Lemma~\ref
{le:martMeasSelection} to find universally measurable kernels
$Q_1(\cdot)$ and $\hat{P}_1(\cdot)$ such that
%
\begin{equation}
\label{eq:QabsCont1} P_1(\cdot) \ll Q_1(\cdot) \ll
\hat{P}_1(\cdot),
\end{equation}
$Q_1(\cdot)$ takes values in $\cQ_1(\cdot)\ \hat{P}_0$-a.s. (and
hence $Q_0$-a.s.), and $\hat{P}_1(\cdot)$ takes values in $\cP
_1(\cdot)\ \hat{P}_0$-a.s. Let
\[
P^1:=P_0\otimes P_1,\qquad  \hat{P}^1:=
\hat{P}_0\otimes\hat{P}_1,\qquad  Q^1:=Q_0
\otimes Q_1.
\]
Then \eqref{eq:QabsCont1} and Fubini's theorem show that $P^1\ll
Q^1\ll\hat{P}^1$ and so we can proceed with $t=2$ as above, using
$\hat{P}^1$ instead of $\hat{P}_0$ as a reference measure, and
continue up to $t=T-1$. We thus find kernels
\[
P_t(\cdot) \ll Q_t(\cdot) \ll\hat{P}_t(
\cdot), \qquad t=0,\dots,T-1,
\]
and we define
\[
Q:=Q_0\otimes\cdots\otimes Q_{T-1}, \qquad \hat{P}:=
\hat{P}_0\otimes \cdots\otimes\hat{P}_{T-1}.
\]
By construction, we have $P\ll Q\ll\hat{P}$ and $\hat{P}\in\cP$;
in particular, $P\ll Q\lll\cP$. Moreover, the fact that $Q_t(\cdot
)\in\cQ_t(\cdot)$ holds $(Q_0\otimes\cdots\otimes Q_{t-1})$-a.s.
and Fubini's theorem yield that $S$ is a generalized martingale under
$Q$ in the sense stated before Lemma~\ref{le:genMartingales}. [We do
not have that $S_t\in L^1(Q)$, which is the missing part of the
martingale property.]
However, Lemma~\ref{le:genMartingales} and Lemma~\ref{le:locMartMeas}
imply that there exists $Q'\sim Q$ under which $S$ is a true
martingale, which completes the proof.
\end{pf*}

\subsection{Superhedging theorem}\label{se:superhedMultiperiod}

We continue with the same setting as in the preceding subsection; that
is, the setup detailed in Section~\ref{se:introMainRes} for the case
when there are no options ($e=0$). Our next aim is to prove the
\hyperref[st]{Superhedging Theorem} in the multi-period market where only stocks are
traded. To avoid some integrability problems in the subsequent section,
we use a slightly smaller set of martingale measures: given a random
variable (``weight function'') $\varphi\geq1$, we define $\cQ
_\varphi:=\{Q\in\cQ\dvtx  E_Q[\varphi]<\infty\}$.

\begin{theorem}\label{th:duality}
Let $\NA(\cP)$ hold, let $\varphi\geq1$ be a random variable and
let $f\dvtx  \Omega\to\R$ be an upper semianalytic function such that
$|f|\leq\varphi$. Then
\[
\sup_{Q\in\cQ_\varphi} E_Q[f] = \pi(f):= \inf \{x
\in\R\dvtx  \exists H\in\cH\mbox{ such that } x+ H\sint S_T\geq f\ \cP\mbox
{-q.s.} \}.
\]
\end{theorem}

We shall give the proof through two key lemmas. First, we provide a
measurable version of the one-step duality (Theorem~\ref
{th:duality1}). Then, we apply this result in a recursive fashion,
where the claim is replaced by a dynamic version of the superhedging
price. The relevance of upper semianalytic functions in this context is
that semianalyticity is preserved through the recursion, whereas
Borel-measurability is not. For the next statement, recall the random
set $\cQ_t$ introduced in Lemma~\ref{le:martMeasSelection}.

\begin{lemma}\label{le:localSuperhedgeSelection}
Let $\NA(\cP)$ hold, let $t\in\{0,\dots,T-1\}$ and let $f\dvtx  \Omega
_{t}\times\Omega_{1}\to\overline{\R}$ be upper semianalytic. Then
\[
\cE_t(f)\dvtx  \Omega_t\to\overline{\R},\qquad  \cE_t(f)
(\omega):= \sup_{Q\in\cQ_t(\omega)} E_Q\bigl[f(\omega,\cdot)
\bigr]
\]
is upper semianalytic. Moreover, there exists a universally measurable
function $y(\cdot)\dvtx  \Omega_t\to\R^d$ such that
%
\begin{equation}
\label{eq:oneStepSuperhedgeLe} \cE_t(f) (\omega) + y(\omega) \Delta S_{t+1}(
\omega,\cdot) \geq f(\omega,\cdot)\qquad  \cP_t(\omega)\mbox{-q.s.}
\end{equation}
for all $\omega\in\Omega_t$ such that $\NA(\cP_t(\omega))$ holds
and $f(\omega,\cdot)>-\infty\ \cP_t(\omega)$-q.s.
\end{lemma}

\begin{pf}
It is easy to see that the mapping $(\omega,Q)\mapsto E_Q[f(\omega,\cdot)]$ is continuous when $f$ is bounded and uniformly continuous.
By a monotone class argument, the same mapping is therefore Borel when
$f$ is bounded and Borel. Using \cite{BertsekasShreve.78}, Proposition~7.26, page
134, and \cite{BertsekasShreve.78}, Proposition~7.48, page
180, it follows that $(\omega,Q)\mapsto
E_Q[f(\omega,\cdot)]$ is upper semianalytic when $f$ is. After
recalling from Lemma~\ref{le:martMeasSelection} that the graph of $\cQ
_t$ is analytic, it follows from the Projection Theorem in the form of
\cite{BertsekasShreve.78}, Proposition~7.47, page 179, that $\cE_t(f)$
is upper semianalytic.

To avoid dealing with too many infinities in what follows, set
\[
\cE_t(f)':=\cE_t(f)\1_{\R}\bigl(
\cE_t(f)\bigr);
\]
then $\cE_t(f)'$ is universally measurable and we have $\cE
_t(f)'(\omega)=\cE_t(f)(\omega)$ unless $\cE_t(f)(\omega)=\pm
\infty$. Consider the random set
\[
\Psi(\omega):=\bigl\{y\in\R^d\dvtx  \cE_t(f)'(
\omega) + y\Delta S_{t+1}(\omega,\cdot) \geq f(\omega,\cdot)
\ \cP_t(\omega)\mbox {-q.s.}\bigr\},\qquad  \omega\in\Omega_t.
\]
We show below that $\{\Psi\neq\emptyset\}$ is universally measurable
and that $\Psi$ admits a universally measurable selector $y(\cdot)$
on that set. Before stating the proof, let us check that this fact
implies the lemma. Indeed, define also $y(\cdot)=0$ on $\{\Psi
=\emptyset\}$. For $\omega$ such that $\Psi(\omega)=\emptyset$,
there are two possible cases. Either $\cE_t(f)'(\omega)<\cE
_t(f)(\omega)$; then necessarily $\cE_t(f)(\omega)=+\infty$, and in
this case \eqref{eq:oneStepSuperhedgeLe} is trivially satisfied for
our choice $y(\omega)=0$. Or, there exists no $y'(\omega)\in\R^d$
solving \eqref{eq:oneStepSuperhedgeLe} with $y(\omega)$ replaced by
$y'(\omega)$; but then
it follows from Theorem~\ref{th:duality1} that we must be in one of
three sub-cases: either $\omega\in N_t$ [i.e., $\NA(\cP_t(\omega))$
fails], or $\{f(\omega,\cdot)=-\infty\}$ is not $\cP_t(\omega
)$-polar, or $\{f(\omega,\cdot)=+\infty\}$ is not $\cP_t(\omega
)$-polar. In the first two sub-cases, the lemma claims nothing. If we
are not in the first two sub-cases and $\{f(\omega,\cdot)=+\infty\}$
is not $\cP_t(\omega)$-polar, it follows via Theorem~\ref{th:FTAP1}
and Lemma~\ref{le:locMartMeas} that $\cE_t(f)(\omega)=+\infty$, so
\eqref{eq:oneStepSuperhedgeLe} is again trivially satisfied for
$y(\omega)=0$.
On the other hand, if $\Psi(\omega)\neq\emptyset$, then our
selector $y(\omega)$ solves \eqref{eq:oneStepSuperhedgeLe} unless
$\cE_t(f)'(\omega)>\cE_t(f)(\omega)$; that is, $\cE_t(f)(\omega
)=-\infty$. However, it follows from Theorem~\ref{th:duality1} that
then again either $\omega\in N_t$ or $\{f(\omega,\cdot)=-\infty\}$
is not $\cP_t(\omega)$-polar.

As a result, it remains to construct a universally measurable selector
for $\Psi$. To this end, it will be necessary to consider the \emph
{difference} $f-\cE_t(f)'$ which, in general, fails to be upper
semianalytic. For that reason, we shall consider a larger class of
functions. Given a Polish space $\Omega'$, recall that we are calling
a function $g$ on $\Omega_t\times\Omega'$ upper semianalytic if for
any $c\in\R$, the set $\{g>c\}$ is analytic, or equivalently, the
nucleus of a Suslin scheme on $\cB(\Omega_t\otimes\Omega')=\cB
(\Omega_t)\otimes\cB(\Omega')$. Let us denote more generally by
$\bA[\cA]$ the set of all nuclei of Suslin schemes on a paving $\cA
$; the mapping $\bA$ is called the Suslin operation. Moreover, let
$\USA[\cA]$ be the set of all $\overline{\R}$-valued functions $g$
such that $\{g>c\} \in\bA[\cA]$ for all $c\in\R$. Hence, if $\cA$
is the Borel $\sigma$-field of a Polish space, $\USA[\cA]$ is the
set of upper semianalytic functions in the classical sense. However, we
shall extend consideration to the class $\USA[\cU(\Omega_t)\otimes
\cB(\Omega')]$, where $\cU(\Omega_t)$ is the universal $\sigma
$-field on $\Omega_t$. It is a convex cone containing both $\USA[\cB
(\Omega_t)\otimes\cB(\Omega')]$ and the linear space of $\cU
(\Omega_t)\otimes\cB(\Omega')$-measurable functions; in particular,
it contains the function $f-\cE_t(f)'$.

On the other hand, the class $\USA[\cU(\Omega_t)\otimes\cB(\Omega
')]$ shares the main benefits of the classical upper semianalytic
functions. This is due to the following fact; it is a
special case of \cite{Leese.75}, Theorem~5.5 (by the argument in the
proof of that theorem's corollary and the subsequent scholium), in
conjunction with the fact that $\bA[\cU(\Omega_t)]=\cU(\Omega_t)$
since $\cU(\Omega_t)$ is universally complete \cite{BertsekasShreve.78}, Proposition~7.42, page 167.

%
\begin{lemma}\label{le:genSelectionThm}
Let $\Gamma\in\bA[\cU(\Omega_t)\otimes\cB(\Omega')]$. Then
$\proj_{\Omega_t} (\Gamma)\in\cU(\Omega_t)$
and there exists a $\cU(\Omega_t)$-measurable mapping $\gamma\dvtx  \proj
_{\Omega_t} (\Gamma) \to\Omega'$ such that\break $\graph(\gamma
)\subseteq\Gamma$.
\end{lemma}

We can now complete the proof of Lemma~\ref
{le:localSuperhedgeSelection} by showing that $\{\Psi\neq\emptyset\}
$ is universally measurable and $\Psi$ admits a universally
measurable selector $y(\cdot)$ on that set.
Fix $y\in\R^d$. Given $\omega\in\Omega_t$, we have $y\in\Psi
(\omega)$ if and only if
\[
\theta_y(\omega,\cdot):= f(\omega,\cdot) - \cE_t(f)'(
\omega) - y\Delta S_{t+1}(\omega,\cdot) \leq0\qquad  \cP_t(\omega)
\mbox{-q.s.}
\]
Moreover, we have $\theta_y\in\USA[\cU(\Omega_t)\otimes\cB
(\Omega_1)]$ as explained above.
Note that for any $P\in\cP_t(\omega)$, the condition $\theta
_y(\omega,\cdot)\leq0\ P$-a.s. holds if and only if
\[
E_{\tilde{P}}\bigl[\theta_y(\omega,\cdot)\bigr]\leq0 \qquad \mbox{for
all } \tilde {P}\ll P,
\]
and by an application of Lemma~\ref{le:integrableProbab}, it is
further equivalent to have this only for $\tilde{P}\ll P$ satisfying
$E_{\tilde{P}}[|\Delta S_{t+1}(\omega,\cdot)|]<\infty$.
Therefore, we introduce the random set
\[
\widetilde{\cP}_t(\omega):= \bigl\{\tilde{P}\in\fP(
\Omega_1)\dvtx  \tilde {P}\lll\cP_t(\omega), E_{\tilde{P}}
\bigl[\bigl|\Delta S_{t+1}(\omega,\cdot )\bigr|\bigr]<\infty\bigr\}.
\]
By the same arguments as at the end of the proof of Lemma~\ref
{le:martMeasSelection}, we can\vspace*{1pt} show that $\widetilde{\cP}_t$ has an
analytic graph (in the classical sense).
Define also
\[
\Theta_y(\omega):= \sup_{\tilde{P}\in\tilde{\cP_t}(\omega)} E_{\tilde{P}}
\bigl[\theta_y(\omega,\cdot)\bigr];
\]
the next step is to show that $\omega\mapsto\Theta_y(\omega)$ is
universally measurable. Indeed,
following the arguments in the very beginning of this proof, the first
term in the difference
\[
E_{\tilde{P}}\bigl[\theta_y(\omega,\cdot)\bigr]=
E_{\tilde{P}}\bigl[f(\omega,\cdot)\bigr]- \cE_t(f)'(
\omega) - y E_{\tilde{P}}\bigl[\Delta S_{t+1}(\omega,\cdot)\bigr]
\]
is an upper semianalytic function of $(\omega,\tilde{P})$; that is,
in $\USA[\cB(\Omega_t)\otimes\cB(\fP(\Omega_1))]$. Moreover, the
second term can be seen as a $\cU(\Omega_t)\otimes\cB(\fP(\Omega
_1))$-measurable function of $(\omega,\tilde{P})$, and the third is
even Borel. As a result,
$(\omega,P)\mapsto\break E_{\tilde{P}}[\theta_y(\omega,\cdot)]$ is in
$\USA[\cU(\Omega_t)\otimes\cB(\fP(\Omega_1))]$. Thus, by the
Projection Theorem in the form of Lemma~\ref{le:genSelectionThm},
\[
\{\Theta_y > c \} = \proj_{\Omega_t} \bigl\{(\omega,\tilde{P})
\in \graph{(\widetilde{\cP}_t)}\dvtx  E_{\tilde{P}}\bigl[
\theta_y(\omega,\cdot )\bigr]>c \bigr\} \in\cU(\Omega_t)
\]
for all $c\in\R$. That is, $\omega\mapsto\Theta_y(\omega)$ is
universally measurable for any fixed $y$.

On the other hand, fix $\omega\in\Omega_t$ and $m\geq1$. Then the
function $y\mapsto\Theta_y(\omega)\wedge m$ is lower semicontinuous
because it is the supremum of the continuous functions
\[
y\mapsto E_{\tilde{P}}\bigl[\theta_y(\omega,\cdot)\bigr]\wedge
m= \bigl(E_{\tilde{P}}\bigl[f(\omega,\cdot)\bigr]- \cE_t(f)'(
\omega) - y E_{\tilde
{P}}\bigl[\Delta S_{t+1}(\omega,\cdot)\bigr]
\bigr)\wedge m
\]
over $\widetilde{\cP}_t(\omega)$. [To be precise, $\Theta_y(\omega
)\wedge m$ may be infinite for some $\omega$. For such $\omega$, the
function $y\mapsto\Theta_y(\omega)\wedge m$ is semicontinuous in
that it is constant.]
By Lemma~\ref{le:semiCaratheodory} below, it follows that $(y,\omega
)\mapsto\Theta_y(\omega)\wedge m$ is $\cB(\R^d)\otimes\cU(\Omega
_t)$-measurable. Passing to the supremum over $m\geq1$, we obtain that
$(y,\omega)\mapsto\Theta_y(\omega)$ is $\cB(\R^d)\otimes\cU
(\Omega_t)$-measurable as well.
As a result, we have that
\[
\graph(\Psi)=\bigl\{(y,\omega)\dvtx  \Theta_y(\omega)\leq0\bigr\} \in\cB
\bigl(\R ^d\bigr)\otimes\cU(\Omega_t).
\]
Now, Lemma~\ref{le:genSelectionThm} yields that $\{\Psi\neq\emptyset
\}\in\cU(\Omega_t)$ and that $\Psi$ admits a $\cU(\Omega
_t)$-measurable selector on that set, which completes the proof of
Lemma~\ref{le:localSuperhedgeSelection}.
\end{pf}

The following statement was used in the preceding proof; it is a slight
generalization of the fact that Carath\'eodory functions are jointly measurable.

%
\begin{lemma}\label{le:semiCaratheodory}
Let $(A,\cA)$ be a measurable space and let $\theta\dvtx  \R^d\times A\to
\overline{\R}$ be a function such that
$\omega\mapsto\theta(y,\omega)$ is $\cA$-measurable for all $y\in
\R^d$ and $y\mapsto\theta(y,\omega)$ is lower semicontinuous for
all $\omega\in A$. Then $\theta$ is $\cB(\R^d)\otimes\cA$-measurable.
\end{lemma}

\begin{pf}
Let $\{y_k\}_{k\geq1}$ be a dense sequence in $\R^d$. For all $c\in
\R$, we have
\[
\{\theta\geq c\} = \bigcap_{n\geq1} \bigcup
_{k\geq1} B_{1/n}(y_k)\times\bigl\{
\theta(y_k,\cdot)> c-1/n\bigr\},
\]
where $B_{1/n}(y_k)$ is the open ball of radius $1/n$ around $y_k$. The
right-hand side is clearly in $\cB(\R^d)\otimes\cA$.
\end{pf}

We now turn to the second key lemma for the proof of Theorem~\ref
{th:duality}. Recall that $\cQ_\varphi=\{Q\in\cQ\dvtx  E_Q[\varphi
]<\infty\}$.

\begin{lemma}\label{le:superhedConstruction}
Let $\NA(\cP)$ hold, let $f\dvtx \Omega\to\R$ be upper semianalytic and
bounded from above and let $\varphi\geq1$ be a random variable such
that $|f|\leq\varphi$. Then there exists $H\in\cH$ such that
\[
\sup_{Q\in\cQ_\varphi} E_Q[f] + H\sint S_T \geq f
\qquad \cP\mbox{-q.s.}
\]
\end{lemma}

\begin{pf}
Let $\cE(f):=(\cE_0 \circ\cdots\circ\cE_{T-1})(f)$; note that the
composition is well defined by Lemma~\ref
{le:localSuperhedgeSelection}. We first show that there exists $H\in
\cH$ such that
%
\begin{equation}
\label{eq:superhedWithcE} \cE(f) + H\sint S_T \geq f\qquad  \cP\mbox{-q.s.}
\end{equation}
Define $\cE^t(f)(\omega):=(\cE_t \circ\cdots\circ\cE
_{T-1})(f)(\omega)$, and let $N$ be the set of all $\omega\in\Omega
$ such that $\NA(\cP_t(\omega))$ fails for some $t\in\{0,\dots,T-1\}$. Then $N$ is a universally measurable, $\cP$-polar set by
Lemma~\ref{le:NAlocal}.
Using Lemma~\ref{le:localSuperhedgeSelection}, there exist universally
measurable functions $y_t\dvtx  \Omega_t\to\R^d$ such that
\[
y_t(\omega) \Delta S_{t+1}(\omega,\cdot) \geq
\cE^{t+1}(f) (\omega,\cdot) -\cE^t(f) (\omega)
\qquad \cP_t(\omega)\mbox{-q.s.}
\]
for all $\omega\in N^c$. Recalling that any $P\in\cP$ is of the form
$P=P_0\otimes\cdots\otimes P_{T-1}$ for some selectors $P_t$ of $\cP
_t$, it follows by Fubini's theorem that
\[
\sum_{t=0}^{T-1} y_t \Delta
S_{t+1} \geq\sum_{t=0}^{T-1} \cE
^{t+1}(f) -\cE^t(f) = f - \cE(f)\qquad  \cP\mbox{-q.s.};
\]
that is, \eqref{eq:superhedWithcE} holds for $H\in\cH$ defined by
$H_{t+1}:=y_t$. Next, we show that
%
\begin{equation}
\label{eq:dpp} \cE(f) \leq\sup_{Q\in\cQ_\varphi} E_Q[f];
\end{equation}
together with \eqref{eq:superhedWithcE}, this will imply the result.
(The reverse inequality of this dynamic programming principle is also
true, but not needed here.)
Recall from below \eqref{eq:tildeXi} that the graph of the random set
\[
\tilde{\Xi}_t(\omega)=\bigl\{(Q,\hat{P})\in\fP(\Omega_1)
\times\fP (\Omega_1)\dvtx  E_Q\bigl[\Delta S_{t+1}(
\omega,\cdot)\bigr]=0, \hat{P}\in\cP _t(\omega), Q\ll\hat{P}\bigr\}
\]
is analytic. We may see its section $\tilde{\Xi}_t(\omega)$ as the
set of controls for the control problem $\sup_{Q\in\cQ_t(\omega)}
E_Q[\cE^{t+1}(f)(\omega,\cdot)]$; indeed, $\cQ_t(\omega)$ is
precisely the projection of $\tilde{\Xi}_t(\omega)$ onto the first
component. As in the proof of Lemma~\ref{le:localSuperhedgeSelection},
we obtain that the reward function $(\omega,Q)\mapsto E_Q[\cE
^{t+1}(f)(\omega,\cdot)]$ is upper semianalytic. Given $\eps>0$, it
then follows from the Jankov--von Neumann Theorem in the form of \cite
{BertsekasShreve.78}, Proposition~7.50, page 184, that there exists a
universally measurable selector $(Q^\eps_t(\cdot),\hat{P}^\eps
_t(\cdot))$ of $\tilde{\Xi}_t(\cdot)$ such that $Q^\eps_t(\omega
)$ is an $\eps$-optimal control; that is,
%
\begin{eqnarray}
\label{eq:QepsOptimizer} E_{Q^\eps_t(\omega)}\bigl[\cE^{t+1}(f) (\omega,\cdot)\bigr]
\geq\sup_{Q\in
\cQ_t(\omega)} E_Q\bigl[\cE^{t+1}(f) (
\omega,\cdot)\bigr]-\eps\nonumber\\[-8pt]\\[-8pt]
\eqntext{\mbox{for all $\omega$ such that }\tilde{
\Xi}_t(\omega)\neq\emptyset.}
\end{eqnarray}
Define also $Q^\eps_t$ and $\hat{P}^\eps_t$ as an arbitrary
(universally measurable) selector of $\cP_t$ on $\{\tilde{\Xi}_t=
\emptyset\}$; note that the latter set is contained in the $\cP
$-polar set $N$ by Theorem~\ref{th:FTAP1}. Now let
\[
Q^\eps:=Q^\eps_0\otimes\cdots\otimes
Q^\eps_{T-1} \quad \mbox{and}\quad  \hat {P}^\eps:=
\hat{P}^\eps_0\otimes\cdots\otimes\hat{P}^\eps_{T-1}.
\]
As in the proof of Theorem~\ref{th:FTAP}, we then have $Q^\eps\ll
\hat{P}\in\cP$ (thus $Q^\eps\lll\cP$), and $S$ is a generalized
martingale under $Q^\eps$. Moreover, recalling that $f$ is bounded
from above, we may apply Fubini's theorem $T$ times to obtain that
\[
\cE(f) = (\cE_0 \circ\cdots\circ\cE_{T-1}) (f) \leq\eps+
E_{Q_0^\eps}\bigl[\cE^1(f)\bigr] \leq\cdots\leq T\eps+
E_{Q^\eps}[f].
\]
By Lemma~\ref{le:genMartingales} and Lemma~\ref{le:locMartMeas},
there exists $Q^\eps_*\in\cQ_\varphi$ such that $E_{Q_*^\eps
}[f]\geq E_{Q^\eps}[f]$. Thus, we have
\[
\cE(f) \leq T\eps+ \sup_{Q\in\cQ_\varphi} E_Q[f].
\]
As $\eps>0$ was arbitrary, we conclude that \eqref{eq:dpp} holds. In
view of \eqref{eq:superhedWithcE}, the proof is complete.
\end{pf}

\begin{pf*}{Proof of Theorem~\ref{th:duality}}
By the same argument as below \eqref
{eq:dualityProofReductionBddAbove}, we may assume that $f$ is bounded
from above. The
inequality ``$\geq$'' then follows from Lemma~\ref
{le:superhedConstruction}, whereas the inequality ``$\leq$'' follows
from Lemma~\ref{le:easyIneq}.
\end{pf*}

\section{The multi-period case with options}\label{se:multiperiodWithOpt}

In this section, we prove the main results stated in the \hyperref[se:intro]{Introduction}.
The setting is as introduced in Section~\ref{se:introMainRes}; in
particular, there are $d$ stocks available for dynamic trading and $e$
options available for static trading. Our argument is based on the
results of the preceding section: if $e=1$, we can apply the already
proved versions of the Fundamental Theorem and the \hyperref[st]{Superhedging Theorem}
to study the option $g^1$; similarly, we shall add the options $g^2,
\dots, g^e$ one after the other in an inductive fashion. To avoid some
technical problems in the proof, we state the main results with an
additional weight function $\varphi\geq1$ that controls the
integrability; this is by no means a restriction and actually yields a
more precise conclusion (Remark~\ref{rk:proofsForMainRes}). We recall
the set $\cQ$ from \eqref{eq:defcQ} as well as the notation $\cQ
_\varphi=\{Q\in\cQ\dvtx  E_Q[\varphi]<\infty\}$.

\begin{theorem}\label{th:mainWeighted}
Let $\varphi\geq1$ be a random variable and suppose that $|g^i|\leq
\varphi$ for $i=1,\dots,e$.
\begin{longlist}[(b)]
\item[(a)] The following are equivalent:
\begin{enumerate}[(ii)]
\item[(i)] $\NA(\cP)$ holds;
\item[(ii)] for all $P\in\cP$ there exists $Q\in\cQ_\varphi$ such
that $P \ll Q$.
\end{enumerate}

\item[(b)] Let $\NA(\cP)$ hold, and let $f\dvtx  \Omega\to\R$ be an
upper semianalytic function such that $|f|\leq\varphi$. Then
\[
\pi(f):=\inf \bigl\{x\in\R\dvtx  \exists (H,h)\in\cH\times\R^e \mbox {
such that } x+ H\sint S_T + hg\geq f\ \cP\mbox{-q.s.} \bigr\}
\]
satisfies
\[
\pi(f)=\sup_{Q\in\cQ_\varphi} E_Q[f] \in(-\infty,\infty],
\]
and there exist $(H,h)\in\cH\times\R^e$ such that $\pi(f)+H\sint
S_T + hg\geq f\ \cP$-q.s.

\item[(c)] Let $\NA(\cP)$ hold, and let $f\dvtx  \Omega\to\R$ be an
upper semianalytic function such that $|f|\leq\varphi$. The following
are equivalent:
\begin{enumerate}[(iii)]
\item[(i)] $f$ is replicable;
\item[(ii)] the mapping $Q\mapsto E_Q[f]\in\R$ is constant on $\cQ
_\varphi$;
\item[(iii)] for all $P\in\cP$ there exists $Q\in\cQ_\varphi$
such that $P\ll Q$ and $E_Q[f]=\pi(f)$.
\end{enumerate}
\end{longlist}
\end{theorem}

\begin{pf}
We proceed by induction; note that if $\NA(\cP)$ holds for the market
with options $g^1,\dots, g^e$, then it also holds for the market with
options $g^1,\dots, g^{e'}$ for any $e'\leq e$.

In the case $e=0$, there are no traded options. Then (a) follows from
Theorem~\ref{th:FTAP} and Lemma~\ref{le:locMartMeas}, and (b) holds
by Theorem~\ref{th:duality}.
To see (c), suppose that (i) holds; that is, $f=x+H\sint S_T\ \cP
$-q.s. for some $x\in\R$ and $H\in\cH$. Using that $|f|\leq\varphi
$, we deduce via Lemma~\ref{le:genMartingales} that $E_Q[f]=x$ for all
$Q\in\cQ_\varphi$, which is (ii). In view of (a), (ii) implies
(iii). Let (iii) hold, let $H\in\cH$ be an optimal superhedging
strategy as in (b), and let $K:=\pi(f)+H\sint S_T-f$. Then $K\geq0\ \cP$-q.s., but in view of (iii) and (a), we have $K=0\ \cP$-q.s.,
showing that $H$ is a replicating strategy.

Next, we assume that the result holds as stated when there are $e\geq
0$ traded options $g^1,\dots, g^e$, and we introduce an additional
(Borel-measurable) option $f\equiv g^{e+1}$ with $|f|\leq\varphi$ in
the market at price $f_0=0$.

We first prove (a). Let $\NA(\cP)$ hold. If $f$ is replicable, we can
reduce to the case with at most $e$ options, so we may assume that $f$
is not replicable.
Let $\pi(f)$ be the superhedging price when the stocks and $g^1,\dots, g^e$ are available for trading (as stated in the theorem); we have
$\pi(f)>-\infty$ by (b) of the induction hypothesis. If $f_0\geq\pi
(f)$, then as $f$ is not replicable, we obtain an arbitrage by
shortselling one unit of $f$ and using an optimal superhedging strategy
for $f$, which exists by (b) of the induction hypothesis. Therefore,
$f_0 < \pi(f)$. Together with (b) of the induction hypothesis, we have
\[
f_0<\pi(f)=\sup_{Q\in\cQ_\varphi} E_Q[f].
\]
As $Q\mapsto E_Q[f]$ is not constant by (c) of the induction hypothesis
and\break $E_Q[|f|]<\infty$ for all $Q\in\cQ_\varphi$, it follows that
there exists $Q_+\in\cQ_\varphi$ such that
\[
f_0 < E_{Q_+}[f] < \pi(f).
\]
A similar argument applied to $-f$ (which is Borel-measurable and not
replicable like $f$) yields $Q_-\in\cQ_\varphi$ such that
%
\begin{equation}
\label{eq:QpmForOpt} -\pi(-f) < E_{Q_-}[f] < f_0 <
E_{Q_+}[f] < \pi(f).
\end{equation}
Now let $P\in\cP$. By (a) of the induction hypothesis, there exists
$Q_0\in\cQ_\varphi$ such that $P\ll Q_0\lll\cP$. By choosing
suitable weights $\lambda_-,\lambda_0,\lambda_+ \in(0,1)$ such that
$\lambda_-+\lambda_0+\lambda_+=1$, we have that
\[
P\ll Q:=\lambda_- Q_- + \lambda_0 Q_0 + \lambda_+ Q_+
\in\cQ\quad  \mbox{and}\quad  E_{Q}[f]=f_0.
\]
This completes the proof that (i) implies (ii). The reverse implication
can be shown as in the proof of Theorem~\ref{th:FTAP}, so (a) is
established. Moreover, the argument for (c) is essentially the same as
in the case $e=0$.

Next, we prove (b). The argument is in the spirit of Theorem~\ref
{th:duality1}; indeed, \eqref{eq:QpmForOpt} states that the price
$f_0$ is in the interior of the set $\{E_Q[f]\dvtx  Q\in\cQ_\varphi\}$
and will thus play the role that the Fundamental Lemma (Lemma~\ref
{le:relInt}) had in the proof of Theorem~\ref{th:duality1}. We
continue to write $f$ for $g^{e+1}$ and let $f'$ be an upper
semianalytic function such that $|f'|\leq\varphi$. Let $\pi'(f')$ be
the superhedging price of $f'$ when the stocks and $g^1,\dots, g^e,
f\equiv g^{e+1}$ are available for trading. Recall that $\pi
'(f')>-\infty$ by Theorem~\ref{th:optimalStrategy} which also yields
the existence of an optimal strategy. Let $\cQ'_\varphi$ be the set
of all martingale measures $Q'$ which satisfy $E_{Q'}[g^i]=0$ for
$i=1,\dots, e+1$ (whereas we continue to write $\cQ_\varphi$ for
those martingale measures which have that property for $i=1,\dots,
e$). Then we need to show that
\[
\pi'\bigl(f'\bigr)=\sup
_{Q'\in\cQ'_\varphi} E_{Q'}\bigl[f'\bigr].
\]
As above, we may assume that $f$ is not replicable given the stocks and
$g^1,\dots, g^e$, as otherwise we can reduce to the case with $e$
options. The inequality
%
\begin{equation}
\label{eq:dualityOptClaimEasyIneq} \pi'\bigl(f'\bigr)\geq\sup
_{Q'\in\cQ'_\varphi} E_{Q'}\bigl[f'\bigr]
\end{equation}
follows from Lemma~\ref{le:easyIneq}; we focus on the reverse inequality
%
\begin{equation}
\label{eq:dualityOptClaimIneq} \pi'\bigl(f'\bigr)\leq\sup
_{Q'\in\cQ'_\varphi} E_{Q'}\bigl[f'\bigr].
\end{equation}
Suppose first that $\pi'(f')<\infty$; then we have $\pi'(f')\in\R
$. In this situation, we claim that
%
\begin{equation}
\label{eq:dualityOptClaimApprox} \mbox{there exist $Q_n\in\cQ_\varphi$ such
that}\qquad  E_{Q_n}[f]\to f_0,\qquad  E_{Q_n}
\bigl[f'\bigr]\to\pi'\bigl(f'\bigr).\hspace*{-31pt}
\end{equation}
Before proving this claim, let us see how it implies \eqref
{eq:dualityOptClaimIneq}.
Indeed, as $f$ is not replicable, there exist measures $Q_\pm\in\cQ
_\varphi$ as in \eqref{eq:QpmForOpt}. Given $Q_n$ from \eqref
{eq:dualityOptClaimApprox}, we can then find weights $\lambda
^n_-,\lambda^n,\lambda^n_+ \in[0,1]$ such that $\lambda^n_-+\lambda
^n+\lambda^n_+=1$ and
\[
Q'_n:=\lambda^n_- Q_- +
\lambda^n Q_n + \lambda^n_+ Q_+ \in\cQ
_\varphi\quad  \mbox{satisfies}\quad  E_{Q'_n}[f]=f_0=0;
\]
that is, $Q'_n\in\cQ'_\varphi$. Moreover, since $E_{Q_n}[f]\to f_0$,
the weights can be chosen such that $\lambda^n_\pm\to0$. Using that
$E_{Q_\pm}[|f'|]<\infty$ as $|f'|\leq\varphi$, we conclude that
\[
E_{Q'_n}\bigl[f'\bigr]\to\pi'
\bigl(f'\bigr),
\]
which indeed implies \eqref{eq:dualityOptClaimIneq}.

We turn to the proof of \eqref{eq:dualityOptClaimApprox}. By a
translation we may assume that $\pi'(f')=0$; thus, if \eqref
{eq:dualityOptClaimApprox} fails, we have
\[
0\notin\overline{\bigl\{E_Q\bigl[\bigl(f,f'\bigr)
\bigr]\dvtx  Q\in\cQ_\varphi\bigr\}}\subseteq\R^2.
\]
Then, there exists a separating vector $(y,z)\in\R^2$ with
$|(y,z)|=1$ such that
\[
0>\sup_{Q\in\cQ_\varphi} E_Q\bigl[yf+zf'\bigr].
\]
But by (b) of the induction hypothesis, we know that
\[
\sup_{Q\in\cQ_\varphi} E_Q\bigl[yf+zf'\bigr] =
\pi\bigl(yf+zf'\bigr).
\]
Moreover, by the definitions of $\pi$ and $\pi'$, we clearly have
$\pi(\psi)\geq\pi'(\psi)$ for any random variable $\psi$.
Finally, the definition of $\pi'$ shows that
$\pi'(yf+\psi)=\pi'(\psi)$, because $f$ is available at price
$f_0=0$ for hedging.
Hence, we have
%
\begin{equation}
\label{eq:ineqPiPrime} 0> \sup_{Q\in\cQ_\varphi} E_Q
\bigl[yf+zf'\bigr] \geq\pi'\bigl(zf'
\bigr),
\end{equation}
which clearly implies $z\neq0$. If $z>0$, the positive homogeneity of
$\pi'$ yields that $\pi'(f')<0$, contradicting our assumption that
$\pi'(f')=0$. Thus, we have $z<0$. To find a contradiction, recall
that we have already shown that there exists some $Q'\in\cQ'_\varphi
\subseteq\cQ_\varphi$. Then, \eqref{eq:ineqPiPrime} yields that $0>
E_{Q'}[yf+zf']=E_{Q'}[zf']$ and hence $E_{Q'}[f']>0=\pi'(f')$ as
$z<0$, which contradicts \eqref{eq:dualityOptClaimEasyIneq}. This
completes the proof of \eqref{eq:dualityOptClaimIneq} in the case $\pi
'(f')<\infty$.

Finally, to obtain \eqref{eq:dualityOptClaimIneq} also in the case
$\pi'(f')=\infty$, we apply the above to $f'\wedge n$ and let $n\to
\infty$, exactly as below \eqref{eq:dualityProofReductionBddAbove}.
This completes the proof of~(b).
\end{pf}

%
\begin{remark}\label{rk:proofsForMainRes}
Given the options $g^1,\dots,g^e$ and $f$, we can always choose the
weight function $\varphi:=1+|g^1|+\cdots+|g^e|+ |f|$; that is, the
presence of $\varphi$ in Theorem~\ref{th:mainWeighted} is not
restrictive. The theorem implies the main results as stated in the
\hyperref[se:intro]{Introduction}. Indeed, the implication from (i) to (ii) in the \hyperref[fft]{First Fundamental Theorem} as stated in Section~\ref{se:introMainRes} follows
immediately from (a), and it is trivial that (ii) implies (ii$^{\prime}$). The
fact that (ii$^{\prime}$) implies (i) is seen as in the proof of Theorem~\ref
{th:FTAP}. The \hyperref[st]{Superhedging Theorem} of Section~\ref{se:introMainRes}
is a direct consequence of (b) and Lemma~\ref{le:easyIneq}. The
equivalence of (i)--(iii) in the \hyperref[sft]{Second Fundamental Theorem} then
follows from (c); in particular, if $\cQ$ is a singleton, the market
is complete. Conversely, if the market is complete, then using $f=\1
_A$, we obtain from (ii) that $Q\mapsto Q(A)$ is constant on $\cQ$ for
every $A\in\cB(\Omega)$. As any probability measure on the universal
$\sigma$-field $\cF$ is determined by its values on $\cB(\Omega)$,
this implies that $\cQ$ is a singleton.
\end{remark}

\section{Nondominated optional decomposition}\label{se:optDecomp}

In this section, we derive a nondominated version of the Optional Decomposition Theorem of \cite{ElKarouiQuenez.95} and \cite
{Kramkov.96}, for the discrete-time case. As in Section~\ref{se:multiperiodWithoutOpt}, we consider the setting introduced in
Section~\ref{se:introMainRes} in the case without options ($e=0$).

\begin{theorem}
Let $\NA(\cP)$ hold, and let $V$ be an adapted process such that
$V_t$ is upper semianalytic and in $L^1(Q)$ for all $Q\in\cQ$ and
$t\in\{1,2,\dots,T\}$. The following are equivalent:
\begin{longlist}[(ii)]
\item[(i)]$V$ is a supermartingale under each $Q\in\cQ$;
\item[(ii)] there exist $H\in\cH$ and an adapted increasing process $K$
with $K_0=0$ such that
\[
V_t=V_0 + H\sint S_t - K_t\qquad  \cP
\mbox{-q.s.}, \qquad t\in\{0,1,\dots,T\}.
\]
\end{longlist}
\end{theorem}

\begin{pf}
It follows from Lemma~\ref{le:genMartingales} that (ii) implies (i).
We show that (i) implies (ii); this proof is similar to the one of
Theorem~\ref{th:duality}, so we shall be brief. Recall the operator
$\cE_t(\cdot)$ from Lemma~\ref{le:localSuperhedgeSelection}. Our
first aim is to show that for every $t\in\{0,1,\dots,T-1\}$, we have
%
\begin{equation}
\label{eq:proofOptDecompDpp} \cE_t(V_{t+1}) \leq V_t \qquad \cP
\mbox{-q.s.}
\end{equation}
Let $Q\in\cQ$ and $\eps>0$. Following the same arguments as in
\eqref{eq:QepsOptimizer}, we can construct a universally measurable
$\eps$-optimizer $Q^\eps_t(\cdot)\in\cQ_t(\cdot)$; that is,
\begin{eqnarray}
E_{Q^\eps_t(\omega)}\bigl[V_{t+1}(\omega,\cdot)\bigr]+\eps\geq
\sup_{Q\in
\cQ_t(\omega)} E_Q\bigl[V_{t+1}(\omega,\cdot)
\bigr] \equiv\cE _t(V_{t+1}) (\omega)\nonumber\\
\eqntext{\mbox{for all $\omega
\in\Omega_t\setminus N_t$},}
\end{eqnarray}
where $N_t=\{\omega\in\Omega_t\dvtx  \NA(\cP_t(\omega)) \mbox{
fails}\}$ is universally measurable and $\cP$-polar (hence $\cQ
$-polar) by Lemma~\ref{le:NAlocal}. Let
\[
Q':=Q|_{\Omega_t} \otimes Q^\eps_t
\otimes Q_{t+1}\otimes\cdots \otimes Q_{T-1},
\]
where $Q_{s}$ is an arbitrary selector of $\cQ_s$ for $s=t+1,\dots,T-1$. Then $S$ is a generalized martingale under $Q'$ and
\[
E_{Q^\eps_t(\omega)}\bigl[V_{t+1}(\omega,\cdot)\bigr] =
E_{Q'}[V_{t+1}|\cF _t](\omega) \qquad \mbox{for
$Q'$-a.e. $\omega$}
\]
by Fubini's theorem, where the conditional expectation is understood in
the generalized sense; cf.\ the \hyperref[se:appendix]{Appendix}. As both sides of this identity
are $\cF_t$-measurable and $Q=Q'$ on $\cF_t$, we also have that
\[
E_{Q^\eps_t(\omega)}\bigl[V_{t+1}(\omega,\cdot)\bigr] =
E_{Q'}[V_{t+1}|\cF _t](\omega)\qquad  \mbox{for $Q$-a.e.
$\omega$.}
\]
Finally, by a conditional version of Lemma~\ref{le:locMartMeas}, there
exists a martingale measure $Q''\in\cQ$ with $Q''=Q'$ on $\cF_t$
such that
\[
E_{Q''}[V_{t+1}|\cF_t]\geq E_{Q'}[V_{t+1}|
\cF_t] \qquad Q\mbox{-a.s.},
\]
and since $V$ is a $Q''$-supermartingale by (i), we also have
\[
V_t\geq E_{Q''}[V_{t+1}|\cF_t] \qquad Q
 \mbox{-a.s.}
\]
Collecting the above (in)equalities, we arrive at
\[
\cE_t(V_{t+1}) - \eps\leq V_t \qquad Q\mbox{-a.s.}
\]
Since $\eps>0$ and $Q\in\cQ$ were arbitrary, we deduce that $\cE
_t(V_{t+1}) \leq V_t\ \cQ$-q.s., and as Theorem~\ref{th:FTAP} shows
that $\cQ$ and $\cP$ have the same polar sets, we have established
\eqref{eq:proofOptDecompDpp}.

For each $t\in\{0,1,\dots,T-1\}$, Lemma~\ref
{le:localSuperhedgeSelection} yields a universally measurable function
$y_t(\cdot)\dvtx  \Omega_t\to\R^d$ such that
\[
\cE_t(f) (\omega) + y_t(\omega) \Delta S_{t+1}(
\omega,\cdot) \geq V_{t+1}(\omega,\cdot)\qquad  \cP_t(\omega)
\mbox{-q.s.}
\]
for all $\omega\in\Omega_t\setminus N_t$. By Fubini's theorem and
\eqref{eq:proofOptDecompDpp}, this implies that
\[
V_t + y_t\Delta S_{t+1}\geq V_{t+1}
\qquad \cP\mbox{-q.s.}
\]
Define $H\in\cH$ by $H_{t+1}:=y_t$; then it follows that
\[
V_t \leq V_0 + H\sint S_t \qquad \cP\mbox{-q.s.},\qquad
t=0,1,\dots,T;
\]
that is, (ii) holds with $K_t:= V_0 + H\sint S_t-V_t$.
\end{pf}

\begin{appendix}
%
\section*{Appendix}\label{se:appendix}

For ease of reference, we collect here some known facts about the
classical (dominated) case that are used in the body of the paper. Let
$(\Omega,\cF,P)$ be a probability space, equipped with a filtration
$(\cF_t)_{t\in\{0,1,\dots,T\}}$. An adapted process $M$ is a \emph
{generalized martingale}
if
\[
E_P[M_{t+1}|\cF_t]=M_t \qquad P
\mbox{-a.s.},\qquad  t=0,\dots,T-1
\]
holds in the sense of generalized conditional expectations, that is,
with the definition
\[
E_P[M_{t+1}|\cF_t]:=\lim_{n\to\infty}
E_P\bigl[M_{t+1}^+\wedge n|\cF_t\bigr] - \lim
_{n\to\infty} E\bigl[M_{t+1}^-\wedge n|\cF_t
\bigr]
\]
and the convention that $\infty-\infty=-\infty$. To wit, if $M_T\in
L^1(P)$, then $M$ is simply a martingale in the usual sense. We refer
to \cite{JacodShiryaev.98} for further background and the proof of the
following fact.

%
\begin{lemmaa}\label{le:genMartingales}
Let $M$ be an adapted process with $M_0=0$. %
The following are equivalent:
\begin{longlist}[(iii)]
\item[(i)]$M$ is a local martingale;
\item[(ii)]$H\sint M$ is a local martingale whenever $H$ is predictable;
\item[(iii)]$M$ is a generalized martingale.
\end{longlist}
If $E_P[M_T^-]<\infty$, these conditions are further equivalent to:
\begin{longlist}[(iv)]
\item[(iv)] $M$ is a martingale.
\end{longlist}
\end{lemmaa}

Let $S$ be an adapted process with values in $\R^d$. The following is
a standard consequence of Lemma~\ref{le:genMartingales}.

%
\begin{lemmaa}\label{le:easyIneq}
Let $f$ and $g=(g^1,\dots,g^e)$ be $\cF$-measurable, and let $Q$ be a
probability measure such that $E_Q[g^i]=0$ for all $i$ and $S-S_0$ is a
local $Q$-martingale. If there exist $x\in\R$ and $(H,h)\in\cH
\times\R^e$ such that
$x+ H\sint S_T + hg \geq f\ Q$-a.s., then $E_Q[f] \leq x$.
\end{lemmaa}

\begin{pf}
If $E_Q[f^-]=\infty$, then $E_Q[f]=-\infty$ by our convention \eqref
{eq:conventionExpectation} and the claim is trivial. Suppose that
$E_Q[f^-]<\infty$; then
$H\sint S_T \geq f-x-hg$ implies $(H\sint S_T)^- \in L^1(Q)$.
Therefore, $H\sint S$ is a $Q$-martingale by Lemma~\ref
{le:genMartingales} and we conclude that $x= E_P[x+H\sint S_T+hg] \geq E_Q[f]$.
\end{pf}

\begin{lemmaa}\label{le:locMartMeas}
Let $Q$ be a probability measure under which $S-S_0$ is a local
martingale, let $\varphi\geq1$ be a random variable and let $f$ be a
random variable satisfying $|f|\leq\varphi$.
There exists a probability measure $Q'\sim Q$ under which $S$ is a
martingale, $E_{Q'}[\varphi]<\infty$, and $E_{Q'}[f]\geq E_{Q}[f]$.
\end{lemmaa}

\begin{pf}
As $Q$ is a local martingale measure, the classical no-arbitrage
condition $\NA(\{Q\})$ holds.
By Lemma~\ref{le:integrableProbab} there exists a probability $P_*\sim
Q$ such that $E_{P_*}[|\varphi|]<\infty$. We shall use $P_*$ as a
reference measure; note that $\NA(\{P_*\})$ is equivalent to $\NA(\{
Q\})$. It suffices to show that
\[
\sup_{Q\in\cQ^*_{\mathrm{loc}}} E_Q[f] \leq\pi(f)\leq\sup
_{Q\in\cQ
^*_\varphi} E_Q[f],
\]
where $\pi(f)$ is the $P_*$-a.s. superhedging price, $\cQ^*_{\mathrm{loc}}$ is
the set of all $Q\sim P_*$ such that $S-S_0$ is a local $Q$-martingale,
and $\cQ^*_{\varphi}$ is the set of all $Q\sim P_*$ such that
$E_Q[\varphi]<\infty$ and $S$ is a $Q$-martingale.
The first inequality follows from Lemma~\ref{le:easyIneq}. The second
inequality corresponds to a version of the classical \hyperref[st]{Superhedging Theorem} with an additional weight function. It can be obtained by
following the classical Kreps--Yan separation argument, where the usual
space $L^1(P)$ is replaced with the weighted space $L^1_{1/\varphi
}(P_*)$ of random variables $X$ such that $E_P[|X/\varphi|]<\infty$.
Indeed, the dual $(L^1_{1/\varphi}(P_*))^*$ with respect to the
pairing $(X,Z)\mapsto E_P[XZ]$ is given by the space of all random
variables $Z$ such that $Z\varphi$ is $P$-a.s. bounded. As a
consequence, the separation argument yields a martingale measure $Q\in
\cQ_\varphi^*$, for if $Z\in(L^1_{1/\varphi}(P_*))^*$ is positive
and normalized such that $E_P[Z]=1$, then the measure $Q$ defined by
$dQ/dP=Z$\vspace*{1pt} satisfies $E_Q[\varphi]=E_P[Z\varphi]<\infty$. The
arguments are known (see, e.g., \cite{Schachermayer.04}, Theorem~4.1),
and so we shall omit the details.
\end{pf}
\end{appendix}

\section*{Acknowledgments}
The authors thank Mathias Beiglb\"ock and Jan Ob{\l}{\'o}j for their
constructive comments.



\printaddresses

\end{document}